\documentclass[preprintnumbers,floatfix,aps,prd,a4paper,nofootinbib,superscriptaddress,eqsecnum,notitlepage]{revtex4-1}

\def\ns{neutron stars }

\def\wrt{with respect to }
\def\a{$\alpha$ }

\def\g{$\gamma$  }
\def\f{$f(R)$  }

\def\eos{equation of state }

\def\qs{quark star }
\def\Mpl{M_{\rm Pl}}

\newcommand{\eq}{\begin{equation}}
\newcommand{\eqe}{\end{equation}}

\newcommand{\eqa}{\begin{eqnarray}}
\newcommand{\eqae}{\end{eqnarray}}

\usepackage{amssymb,epsf}
\usepackage{latexsym}
\usepackage{hyperref,natbib,textcase,bm}
\usepackage{amsmath,mathrsfs}
\usepackage{graphicx,epsfig}
\usepackage[utf8x]{inputenc}
\usepackage{graphics}
\usepackage[english]{babel}
\usepackage{subfig}
\usepackage{capt-of}
\usepackage{array}
\usepackage{caption}
\usepackage{color}
\usepackage[fit]{truncate}
\usepackage{comment} 
\usepackage{float}
\usepackage[margin=2cm,a4paper]{geometry} 
\usepackage{caption}
\usepackage{multirow}
\usepackage{epstopdf}

\begin{document}

\title{Modified gravity with logarithmic curvature corrections and the structure of relativistic stars}
\author{Hamzeh Alavirad \footnote{hamzeh.alavirad@kit.edu}, Joel M. Weller \footnote{joel.weller@kit.edu}}
\address{Institute for Theoretical Physics, Karlsruhe Institute of Technology, 76128 Karlsruhe, Germany}

\begin{abstract}
We consider the effect of a 
logarithmic $f(R)$ theory,
motivated by the form of the one-loop effective action arising from 
gluons in curved spacetime, on the structure of relativistic stars.
In addition to analysing the consistency constraints on the potential
of the scalar degree of freedom, we discuss the
possibility of observational features arising from
a fifth force in the vicinity of the neutron star surface.
We find that the model exhibits a chameleon effect that
completely suppresses the effect of the modification on scales
exceeding a few radii, but 
close to the surface of the neutron star, the deviation from General Relativity can
significantly affect the surface redshift that 
determines the shift in absorption (or emission) lines.  
We also use
the method of perturbative constraints to
solve the modified Tolman-Oppenheimer-Volkov equations
for normal and self-bound neutron stars (quark stars).
\end{abstract}

\preprint{KA-TP-20-2013}

\maketitle

\section{Introduction}
\par The modification of the Einstein-Hilbert (EH) action to include higher 
order curvature invariants has a distinguished history,
beginning just a few years after the introduction of General Relativity (GR)
\cite{Edd, *Weyl}.
However, it was the realisation that renormalization 
at one loop demands that the EH action
be supplemented with higher order terms that
 stimulated interest in modifications in the strong gravity regime,
 such as Starobinsky's well-known curvature
  driven inflationary scenario \cite{starobinsky}.
The possibility that such corrections could affect gravitational
phenomenology at low energies was not seriously considered until the
discovery of the acceleration of the 
expansion of the universe \cite{Riess:1998cb, *Perlmutter:1998np},
whereupon $f(R)$ models in particular, in which the EH action is replaced with
a more general function of the Ricci scalar, have been intensely studied
by many authors (see \cite{soti,DeFelice:2010aj} for comprehensive reviews).

Modifications of gravity that lead to deviations in the
low energy regime, corresponding to the late universe, must,
in addition to compatibility with cosmological observations
and internal consistency requirements,
stand up to a host of constraints
arising from equivalence principle tests and solar system measurements on local scales.
Since $f(R)$ theories can be reformulated as a scalar-tensor
theory with a fixed coupling to matter, these tests are sufficient to
rule out the models, unless the fifth force generated by the scalar degree of freedom
is effectively screened, as in the chameleon mechanism \cite{Khoury:2003aq,*Khoury:2003rn,Brax:2008hh}.

By comparison, the strong gravity regime
is poorly constrained by observations \cite{PsaltisReview}.
One can consider the stability of relativistic
stars in $f(R)$ gravity as a test of the theory's viability; indeed, it was
claimed by the authors of \cite{kobay} that the
formation of compact objects is actually prohibited in
 cosmologically successful $f(R)$ models that modify the EH action in the low-curvature
regime, due to the presence of a physically accessible curvature singularity.
However, it was later shown explicitly that this claim does not hold, and that by
taking account of the chameleon effect (i.e. considering the nonlinearity of 
the field equations \cite{upad})  or using a more realistic equation of state  \cite{babi, babi2}
such solutions can be constructed.

One difficulty with the $f(R)$ models discussed in the last paragraph
is that the purported instabilities occur when treating the model as exact at scales far removed
from the phenomena they were constructed to describe.
This consideration has led Cooney et al. \cite{cooney}. to treat relativistic stars as a
framework in which to study $f(R)$ models under the 
assumption that the modifications are next to leading order corrections to the EH action. 
Using the method of perturbative constraints and corrections of the form $R^{n+1}$, 
they showed that the predicted mass-radius relation for neutron stars differs from that calculated in
the General Relativity, although this is degenerate with the neutron star equation of state.
Subsequent studies by other authors have focused on $R$-squared models with
$f(R)=R+\alpha R^2$ \cite{Arapoglu,Orellana:2013gn} and also $R^{\mu\nu}R_{\mu\nu}$ 
\cite{deli} terms (see also \cite{Santos:2011ye,Kamiab:2011am})
where in the former the
value of $\alpha$ is constrained 
to be $\alpha\lesssim 10^{6}{\rm m}^2$ (cf. \cite{PhysRevD.81.064030} for a 
detailed discussion on this point.)
Recently, the same 
$f(R)$ model was applied to a neutron star with a strong magnetic field and 
the constraints on the parameter $\alpha$ obtained as $\alpha\le 10^{5}{\rm m}^2$ \cite{Cheoun:2013tsa}. The problem of gravitational collapse and hydrostatic equilibrium in $f(R)$ gravity has also been
considered by several authors \cite{Capozziello:2011nr,*Capozziello:2011gm,*Bamba:2011sm,*Borisov:2011fu,*Reverberi:2012ew,*Sharif:2012zzb}.

\par In this paper, by considering the semiclassical approach to 
quantum gravity, we propose a phenomenological 
 $f(R)$ model of the form $R+\alpha R^2
+\beta R^2\ln(R/\mu^2)$ that is relevant for the strong field regime in the interior 
of relativistic stars.
$f(R)$ theories with logarithmic terms have been previously considered as models of dark energy
\cite{Nojiri:2003ni} and modified gravity models of this form have also been discussed in early works
\cite{baib, *vilenkin,*Shore1980}
in the context of the Starobinsky inflationary model.
Cosmological evolution in a logarithmic model arising from a running gravitational coupling
has also been studied in the recent work \cite{Guo:2013swa}.

It is well known that 
in the absence of a viable theory of quantum gravity, 
semiclassical methods like 
quantum field theory in curved spacetime 
 are useful tools to study 
the influence of gravitational fields on quantum phenomena \cite{birrell}. 
The curvature of spacetime 
modifies the gluon propagator with terms proportional to the Ricci scalar in 
a constant-curvature spacetime locally around the gluons.
As was first shown by Leen \cite{Leen1983417} and Calzetta et al. 
\cite{PhysRevLett.55.1241,*PhysRevD.33.953} (see also \cite{PhysRevD.34.3025}),
 one-loop renormalization of non-Abelian gauge theories in a general curved
spacetime induces terms logarithmic in $R$ that dominate at large curvature. 
Neutron stars probe the 
dense QCD phase diagram at low temperature and high 
baryon densities, where the baryon density in 
the stellar interior can reach an order
of magnitude beyond the nuclear saturation density
$\rho_{ns}=2.7\times 10^{17}\textmd{kg m}^{-3}$.
In such a dense medium, where the
strong nuclear force plays an paramount role,
we
consider the effect of 
corrections to the EH action involving terms of 
the form $\alpha R^2+\beta R^2\ln(R/\mu^2)$ on the observational features
of the neutron star. 

\par 
We shall also consider the effect of the $f(R)$ model
on a separate class of neutron stars:
self-bound stars, consisting of strange quark matter
with finite density but zero pressure at their surface \cite{1970PThPh..44..291I, witten,Alcock:1988re}.
The interior of the star is made up of deconfined quarks that form a colour superconductor,
leading to a softer equation of state
with possible observable effects on the minimum mass, radii, 
cooling behaviour and other observables \cite{Weber:2006iw,*Lattimer:2000nx,*Ozel:2006bv}.

The structure of this paper is as follows. 
In Sec.  \ref{MGPsec} we motivate the $f(R)$ model by considering the
calculation of the gauge invariant effective action for gauge fields
in curved spacetime.
Then in section \ref{cons}, we investigate constraints imposed upon the
model from the requirements of internal consistency and compatibility
with observations, and discuss the potential observational signatures
due to a change in the effective gravitational constant near the surface of the star.
In section \ref{NS} the structure of relativistic stars is considered in the framework of the $f(R)$ theory,  and we summarise our results in section \ref{summary}. 
Unless otherwise stated, we use a metric with signature +2, and
define the Riemann tensor by
$R^\epsilon{}_{\sigma\mu\nu} = \partial_\mu\Gamma^\epsilon{}_{\nu\sigma}
    - \partial_\nu\Gamma^\epsilon{}_{\mu\sigma}
    + \Gamma^\epsilon{}_{\mu\lambda}\Gamma^\lambda{}_{\nu\sigma}
    - \Gamma^\epsilon{}_{\nu\lambda}\Gamma^\lambda{}_{\mu\sigma}$. We use units 
    such that $\hbar=c=1$.

\section{Motivations}\label{MGPsec}
The behaviour of gauge theories in curved spacetime
was studied in detail by several authors  some thirty years ago,
with the intention of seeing if quantitatively new effects
appear in the high-curvature limit (cf. \cite{ParkerToms} for a textbook discussion and
original references). 
In particular it was shown by Calzetta et al. \cite{PhysRevLett.55.1241,*PhysRevD.33.953}
that
for a pure gauge theory in a general curved space-time, the effective value of the gauge coupling constant can become small in the high curvature limit,
due to the presence of $\ln(R/\mu^2)$ terms in the renormalised gauge-invariant effective action:
a situation referred to as curvature-induced asymptotic freedom. 
Without going into details, in this section we sketch how this result comes about, and use the
form of the 
full result to motivate the phenomenological $f(R)$ theory that will be investigated in more
detail in the remainder of the paper. 

The classical action for a pure gauge field is\footnote{
In this section we use the shorthand $(f,g)=\int d^dx \sqrt{-g} f_a(x)g_a(x)$ for fields $f$, $g$ with components
$f_a$, $g_a$.
} $S[A] = -\tfrac{1}{4}(F_{\mu\nu},F^{\mu\nu})$, where $A_\mu = A_{\mu, a} t_a^{\rm adj}$
is a gauge field in the adjoint representation, $[t_a^{\rm adj},t_b^{\rm adj}] = if_{abc}t_a^{\rm adj}$,  and the field strength is
\eq
F_{\mu\nu,a} = \nabla_\mu  A_{\nu, a} -  \nabla_\nu  A_{\mu, a} +e_g f_{abc} A_{\mu, b} A_{\nu, c},
\eqe
in terms of the metric covariant derivative $\nabla_\mu$.
The generating function
for disconnected graphs in the presence 
of a background gauge field $A_\mu$ and a source $J_\mu$ is
\eq
Z[J,A] = \int \mathcal{D}[a]\mathcal{D}[\eta]\mathcal{D}[\bar\eta] \exp\left( i\left[ S[A+a] + S_{\rm gf} +S_{\rm ghost} +S_{\rm grav}
+ (J_\mu,a^\mu) \right] \right),
\eqe
where $S_{\rm gf}= -\tfrac{1}{2\omega}(D\cdot a, D\cdot a,)$ is the gauge fixing term
and $S_{\rm ghost} = -\int d^dx\sqrt{-g} \bar\eta D\cdot(D+a)\eta$
is the ghost field action.
Here $D$ refers to the (gauge) covariant derivative $D_\mu = \nabla_\mu+ie_g A_\mu$.
Renormalizability in curved spacetime requires the inclusion of squared-curvature terms
in addition to the Einstein-Hilbert action
\eq
S_{\rm grav} = \int d^d x \sqrt{-g} (-\Mpl^2\Lambda + \tfrac{\Mpl^2}{2} R +\alpha_1R^{\mu\nu\rho\sigma}R_{\mu\nu\rho\sigma} +\alpha_2 R^{\mu\nu}R_{\mu\nu}
+\alpha_3 R^2),
\eqe
where $\Mpl^2=1/8\pi G$ and the authors of \cite{PhysRevLett.55.1241,*PhysRevD.33.953}
use a metric with signature -2 and, relative to our convention, the opposite sign for $R^\epsilon{}_{\sigma\mu\nu}$.
The gauge-invariant effective action $\Gamma[A]$ is obtained
via a Legendre transformation from the functional $W=-i\ln(Z)$.
To one-loop order, it is given by 
\eq
\Gamma[A] = S[A] + S_{\rm grav}+ \tfrac{i}{2}\ln\det ( K) -i\ln \det (D^2),
\eqe
where
\eq \label{CalzettaK}
 K_{\mu\nu} = g_{\mu\nu} D^2 - (1-1/\omega)D_\mu D_\nu -2i e_g F_{\mu\nu}+R_{\mu\nu},
\eqe
and $D^2 = D_\mu D^\mu$. Since $\Gamma[A]$ is gauge invariant, the calculation may be
simplified without affecting the final result by choosing the Feynman gauge $\omega=1$.
In general, one has a choice concerning the separation
of the full action into a free part and an interacting part, which determines which terms provide
propagators entering into Feynman diagrams and which provide vertices. The
above choice corresponds to taking the free part to consist of all terms quadratic in the quantum fields
$a, \bar\eta, \eta$.\footnote{
Another possibility is to treat terms involving the background field $A$
as interaction terms,
in which case the inverse propagator involves only the first and last terms in 
(\ref{CalzettaK}).
As shown in \cite{PhysRevD.33.953}, the final results for the two methods agree.
}
Regularising using dimensional regularisation gives
\begin{multline}
\Gamma[A] =  S[A_B] + S_{\rm grav,B}
+\frac{1}{(4\pi)^{d/2}}\int d^dx \sqrt{-g} \frac{1}{(-R/6)^{2-d/2}}
\bigg\lbrace
\left[1+\frac{1}{12}\left(1-\frac{d}{2} \right)\right]\Gamma(2-\tfrac{d}{2})Ce_g^2 \mu^{(4-d)} F_{\mu\nu,a}F^{\mu\nu}_a+  \\
+\Gamma(2-\tfrac{d}{2})N\left[ -\frac{1}{9}\frac{(d+1)}{d(d-2)}R^2 + \frac{d-17}{360}R_{\mu\nu\rho\sigma} 
R^{\mu\nu\rho\sigma} + \frac{92-d}{360}R_{\mu\nu}R^{\mu\nu}\right]+
\sum_{j=3}^\infty \frac{\Gamma(j-\tfrac{d}{2})}{(-R/6)^{j-2}}{\rm tr}[H_{j}]\bigg\rbrace,
\end{multline}
where $\delta_{ab}C=tr(t_a^{\rm adj},t_b^{\rm adj})$, $N$ is the dimension of the gauge group
and
$H_{j}$ stands for curvature and field strength terms entering into the relevant Schwinger-DeWitt series. The subscript $B$ indicates that these terms involve bare quantities.
Adopting the minimal subtraction scheme, the renormalised 
gauge-invariant effective action $\Gamma[A]$ is found to be
\eq
\begin{split}
\Gamma[A] = & \  S[A] + S_{\rm grav} - \\
& -\frac{1}{16\pi^2}\int d^4 x\sqrt{-g}
\left[ \ln\left( \frac{-R/6}{4\pi\mu^2} \right) +\gamma_E \right]
\left[ \tfrac{11}{12}e_g^2 C F_a^{\mu\nu}F_{\mu\nu,a} + (-\tfrac{13}{360}R_{\mu\nu\rho\sigma} 
R^{\mu\nu\rho\sigma} 
+\tfrac{11}{45}R_{\mu\nu}R^{\mu\nu} - \tfrac{5}{72}R^2
 )N \right],
\end{split}
\eqe
where $S[A]$ + $S_{\rm grav} $ contain finite renormalised coefficients and  $\gamma_E$ is the Euler-Mascheroni constant.
Here, the minus sign is kept in the logarithm to emphasise that it is $-R/6$ that plays the role of
`squared mass' in the loop integrals, 
however, the integrals leading to this result are well-defined regardless of the sign of $R$ \cite{PhysRevD.33.953}. 
From a phenomenological
perspective the $\ln(-1) = i\pi$ is simply another finite contribution entering into the coefficients of the squared curvature and field strength terms in the gravitational and gauge field actions.
It should also be noted that for effects such as
curvature-induced asymptotic freedom, only the real part $\ln(|R|/|R_0|)$,
where $R_0$ is a scalar curvature chosen so that $e_g$ is small and so perturbation theory is valid,
enters the expressions for the effective coupling constant $e_g^{\rm eff}$  \cite{PhysRevLett.55.1241}. 

In a maximally symmetric spacetime with constant curvature, the gravitational part of the effective Lagrangian for
a non-Abelian gauge field such as the gluon field
would thus consist of $R^2$ and $R^2\ln(R/\mu^2)$ terms.
On large scales, 
far removed from those relevant for subatomic particles,
relaxing the constant curvature condition would lead to a
non-standard dependence of the gravitational action on the curvature.
In this article we are interested in the effect of modifications to
the EH action on the structure of relativistic stars, where QCD plays an important role.
Motivated by the results summarised in this section, we propose a
phenomenological $f(R)$ model
\begin{equation}
 \mathcal{S}_{tot}=\frac{M^2_{Pl}}{2}\int d^4x\sqrt{-g} \left[R+\alpha R^2
+\beta R^2\ln(R/\mu^2)\right]+\mathcal{S}_{m}\label{f(R)},
\end{equation}
where the constants $G$, $\alpha$ and $\beta$ should be determined by observations. 
As we consider only astrophysical scales, we do not include the effect of the 
cosmological constant term.
We note that modified gravity theories of this form have also been discussed in early works
discussing the effective gravitational action of conformally covariant fields
\cite{baib, *vilenkin,*Shore1980} in the context of the Starobinsky inflationary model.

As we are considering neutron stars, a natural choice of the parameter 
$\mu$ should contain the relevant mass scales. 
We will assume 
\eq\label{muDef}
\mu=m_n^2/M_{Pl},
\eqe 
where $m_n$ is the neutron mass and $M_{Pl}$ is the Planck mass.
$\mu^2$ is then
of the order of the curvature of a typical neutron star. 

\section{Constraints on the model}\label{cons}

In section \ref{NS} we shall investigate
the phenomenology of relativistic stars in the $f(R)$ theory described by the action (\ref{f(R)}),
working in the metric formalism.
Firstly, in sections \ref{SubSec:Consistency} and  \ref{SubSec:Observations} we
consider consistency and observational 
constraints to check the viability of the model in 
such a medium. 
It is important to emphasise that we treat
the model as an effective theory valid in the interior and vicinity of ultra-dense matter,
and so do not consider cosmological or solar system tests. 

\subsection{Consistency constraints}\label{SubSec:Consistency}
An $f(R)$ model inevitably introduces a scalar degree of freedom,
which is constrained by the requirement that the model must be free
of instabilities \cite{soti}.
Such consistency constraints are not always obvious at first sight; indeed,
generalising the findings of Dolgov and Kawasaki \cite{Dolgov:2003px},
 it was
pointed out by Frolov \cite{frolov} 
that many $f(R)$ models that deviate from General Relativity in the
infrared possess a crippling nonlinear instability. In this section, we illustrate
how these constraints can restrict the parameters of our model. 

\begin{figure}[th!]
\centering
	\begin{tabular}{ccc}
        
        \epsfig{file=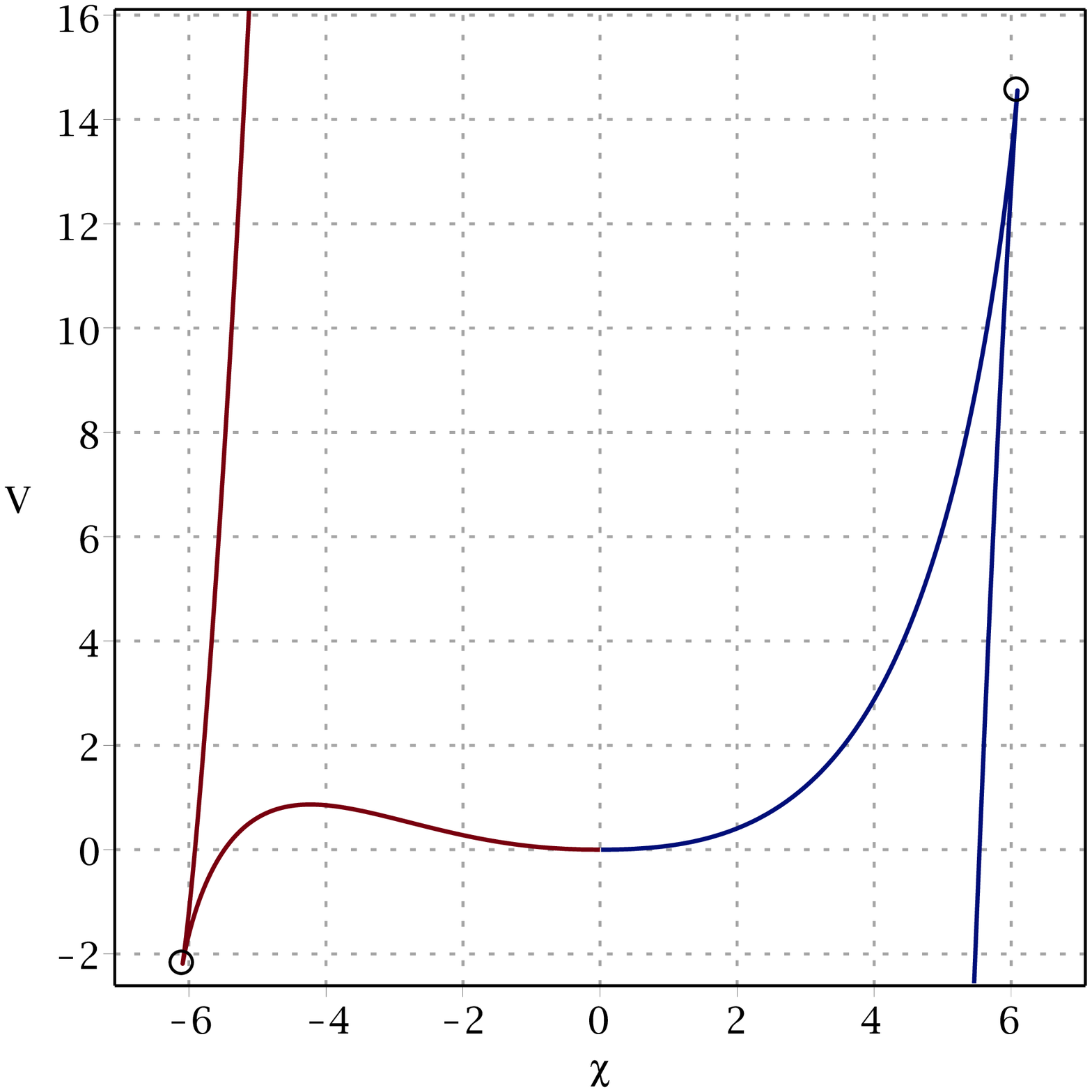, width=0.3\linewidth} &
        \epsfig{file=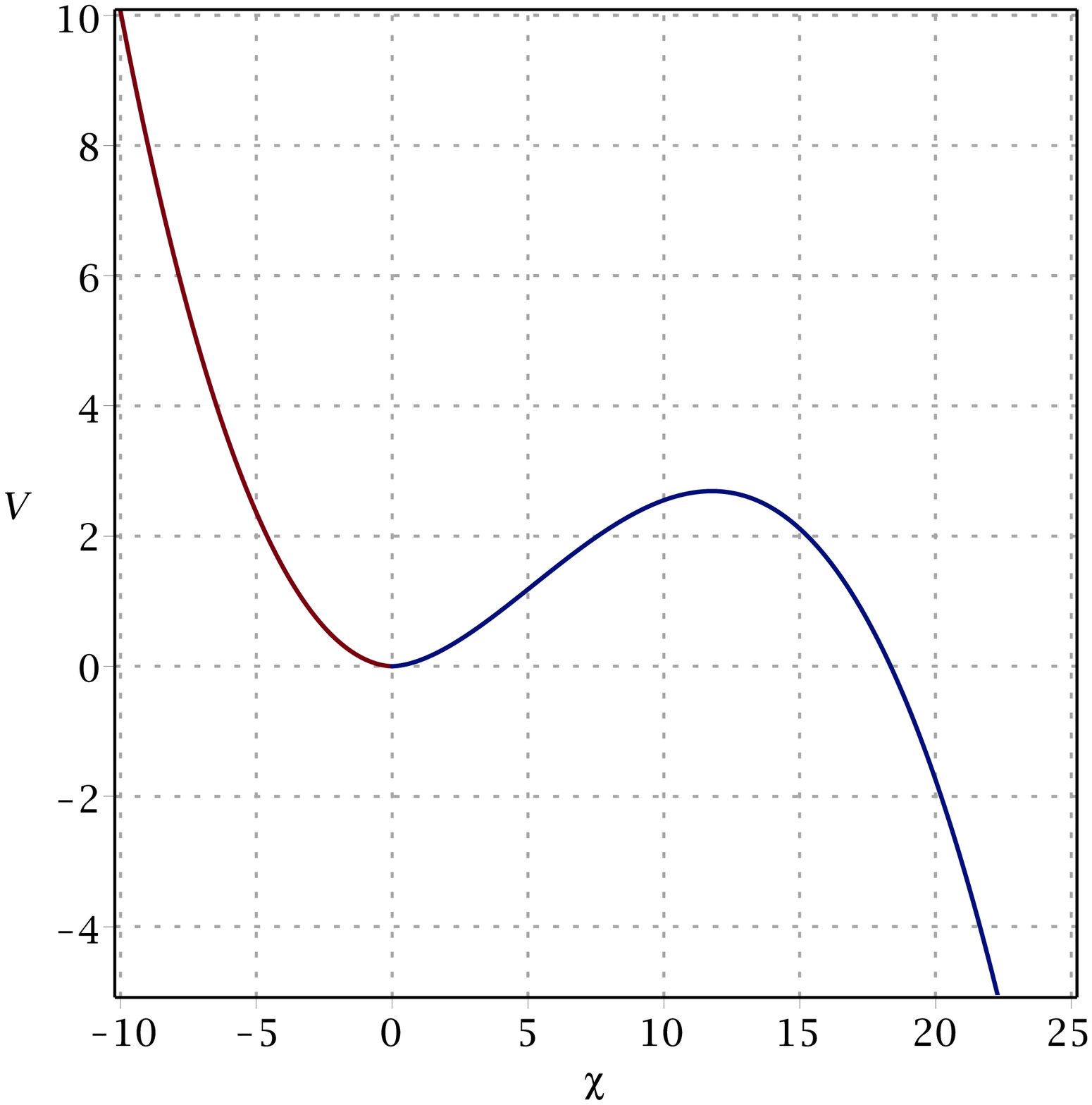, width=0.3\linewidth} &
        \epsfig{file=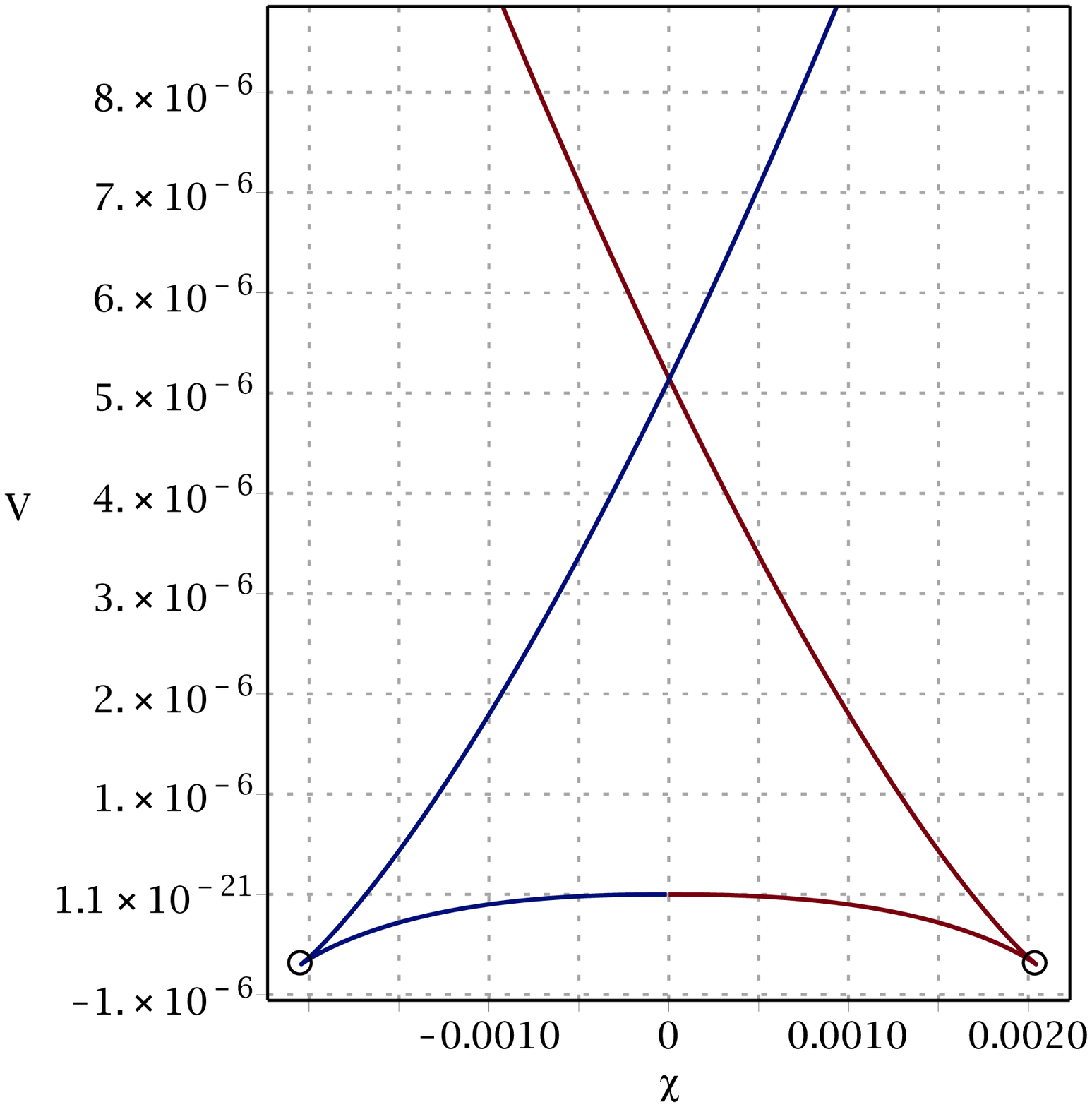, width=0.3\linewidth}\\
	\end{tabular}

        \caption{(Color online.)
        The potential $V(\chi)$ corresponding 
        to positive (blue) and negative (red) $R$. The branch points
        at $\chi = \chi_*$ are indicated by the black circles.
        Large values, $\alpha=\mu=1$, $|\beta|=0.25$ have been chosen to illustrate the important features.
        Left panel: Negative $\beta$.
        Middle panel: Positive $\beta$. The apparent minimum at $\chi=0$ in the middle panel is actually a maximum
        with branch points at $\chi=\chi_* \ll 1$, as can be seen in the right panel, which is a close-up
        of the region around $\chi=0$ for $\beta>0$.
        }\label{Cons_beta_plots}
\end{figure}

From (\ref{f(R)}) we have
\eq
  f(R)= R+\alpha R^2 +\beta R^2\ln\frac{R}{\mu^2}.
  \label{f(R)4}
\eqe
In this section and throughout this paper, we shall restrict ourselves to the case in which the $R^2\ln(R/\mu^2)$
term is subdominant to the $R^2$ term i.e. $|\gamma| \ll 1$, where
\eq
\gamma \equiv \beta/\alpha.
\eqe
The system is best studied in the original frame (i.e without performing a conformal
transformation to the Einstein frame). The equation of motion for the scalar degree of freedom is
\eq
\Box f_{R}  = \frac{2f-f_{R}R}{3} +\frac{8\pi G}{3}T,
\eqe 
where $T$ is the trace of the stress-energy tensor.
Defining
\eq
\chi\equiv f_{R}-1\;,
\eqe 
 where $f_R\equiv df(R)/dR$, this can be recast in the form
\eq
\Box \chi = \frac{dV}{d\chi} - \mathcal{F},
\label{chi_EOM}
\eqe
where $\mathcal{F} = -(8\pi G/3)T$ appears as a force term and $V$ is a potential satisfying
\eq
 \frac{dV}{d\chi} = \frac{1}{3}(2f-f_{R}R)
\eqe
In the model at hand, the form of $f(R)$ and its derivatives are given by
\eqa
f(R) &=& R + \alpha R^2 +\beta R^2 \ln (R/\mu^2), \\
f_{R}(R) &=& 1+(2\alpha+\beta) R +2\beta R \ln (R/\mu^2), \\
f_{RR}(R) &=& 2\alpha+3\beta +2\beta \ln (R/\mu^2), 
\label{fRR_eq}
\eqae
so that
\eq
 \frac{dV}{d\chi} = \frac{1}{3}(R-\beta R^2).
 \label{dVdchi_eq}
\eqe
As we shall see in Sec. \ref{NS}, the modified Einstein equations involve $f_{RR}$, which is 
not analytic at $R=0$. Hence, we shall restrict our analysis to non-negative values of the curvature scalar. 
To obtain the form of the potential without inverting, one can multiply (\ref{dVdchi_eq}) 
by (\ref{fRR_eq}) and integrate with respect to $R$ to yield the parametric equations\footnote{
Note that in order to show the full form of the potential obtained from (\ref{f(R)4}) using the range $R\in (-\infty,\infty)$, 
we have adjusted the numerical factors here so that the arguments of the logs depend on $R^2$. 
We shall only consider the part corresponding to $R\ge 0$.
}
\eq
\chi(R) = R\left[ 2\alpha+\beta +\beta \ln\left( \frac{R^2}{\mu^4} \right) \right],
\eqe
and
\eq
V(R) = -\frac{R^2}{9}\left\lbrace \beta R \left[2\alpha+\tfrac{7}{3}\beta + \beta \ln\left( \frac{R^2}{\mu^4} \right)\right]  
-3\alpha -3\beta -\tfrac{3}{2}\beta \ln\left( \frac{R^2}{\mu^4} \right) \right\rbrace.
\eqe
The potential is shown in Fig. \ref{Cons_beta_plots}.
One can see immediately that in the limit of large curvature ($R\rightarrow \infty$) $V\rightarrow -\infty$
while $\chi\rightarrow \operatorname{sgn}(\beta) \infty$ (for negative $\beta$ the potential turns back on 
itself after an inflection point to reach negative $\chi$.) This should be contrasted with the behavior of the basic 
$f(R)=R+\alpha R^2$ model, where the potential is a simple quadratic in the $\chi$-field. Thus, 
Frolov's singularity --- in which the curvature singularity is a finite distance in field 
and energy values away from the stable solution --- will be avoided.

What is the the nature of the stable solution in this model in the absence of matter?
From (\ref{dVdchi_eq}) we note that there are two stationary points, at $R=0$ and $R = 1/\beta$ 
respectively; 
to ensure perturbative stability, the scalar degree of freedom should satisfy 
the important requirement that its squared mass term 
is positive $m_\chi^2 \equiv d^2V /d\chi^2 >0$.
 It follows from (\ref{fRR_eq}) that 
\eq
m_\chi^2(R)  = \frac{dR}{d\chi}\frac{d}{dR}\left( \frac{2f-f_{R}R}{3} \right)  = \frac{1-2\beta R}{3f_{RR}},
\label{m^2_Def}
\eqe
however, one cannot substitute $R=0$ into this expression due to the singularity in the $\ln$ term in (\ref{fRR_eq}).
For small $\epsilon$ we have from the form of the potential 
\eq
V(R=\pm \epsilon) =\frac{\alpha}{3}[1+\gamma+\gamma\ln(\epsilon/\mu^2)]\epsilon^2+\mathcal{O}(\epsilon^3),
\eqe
which should be positive as $\epsilon\rightarrow 0$ if $R=0$ is a minimum.
Assuming $|\gamma|\ll 1$, this is true only when $\beta<0$, regardless of the sign of $\alpha$.

For $R = 1/\beta$ to be a minimum, one needs $f_{RR}(R=1/\beta)<0$. 
As we do not consider negative curvature, $\beta>0$ and the condition is 
equivalent to
\eq
R_*\beta >1,
\eqe
where we have defined
\eq
R_* = \mu^2 \exp\left(  -\tfrac{3}{2}-\gamma^{-1}  \right).\label{Rstar}
\eqe
When
$|\gamma| \ll 1$, the dimensionless ratio $R_*/\mu^2$ is exponentially large for 
negative $\gamma$ and exponentially small for positive $\gamma$.
We conclude that the stationary point at $R=1/\beta$ is only stable for negative alpha. 

Since maximally symmetric solutions lead to a constant Ricci scalar [and so the derivatives of 
$\chi$ vanish in (\ref{chi_EOM})] one can conclude from this that the maximally symmetric solution is Minkowski spacetime ($R=0$) when $\beta<0$ and de Sitter spacetime when $\beta>0$, $\alpha<0$.


We can also analyse the sign of $m_\chi^2$ away from the stationary points.
For negative $\beta$ we find
\eq
m^2_\chi >0 \quad \Rightarrow \quad R<R_*  \qquad\qquad (\beta<0),
\label{m^2cond}
\eqe
which in terms of $\chi$ is $\chi < \chi_* \equiv -2\beta R_*$. 
For positive $\beta$ one must also take the numerator of (\ref{m^2_Def}) into account, giving
\eq \label{PostbetaNumCases}
m^2_\chi >0 \quad \Rightarrow \quad  \bigg\lbrace \begin{array}{cc}  R_* < R< \frac{1}{2\beta}, & R_*<\tfrac{1}{2\beta} \\ 
R_* > R > \frac{1}{2\beta}, & R_* > \tfrac{1}{2\beta} \end{array}
  \qquad\qquad (\beta>0).
\eqe
The relevant interval depends on whether the condition
$R_*<\tfrac{1}{2\beta}$ is satisfied. Since we are only interested in positive $\beta$
here we can write this as
\eq
 e^{\gamma^{-1}-\ln|\gamma|} > 2e^{-3/2} |\mu^2 \alpha|.
\eqe
As discussed in Sec. \ref{NS}, in order to make use of the method of perturbative constraints
we shall work with parameter values such that $|\alpha \mu^2|\ll 1$.
Hence, when $|\gamma| \ll 1$,  $R_*<\tfrac{1}{2\beta}$
is easily satisfied if $\alpha>0$. 
Similarly, $R_*> \tfrac{1}{2\beta}$ when $\alpha<0$.

The requirement that the graviton is not a ghost\footnote{
As calculated by expanding the propagator about Minkowski spacetime.
}, or equivalently that the effective gravitational constant 
$G_{\rm eff}$
is positive, imposes the well-known condition $f_{R}(R)>0$. Using the definition of $\chi$ this gives $\chi>-1$. We can write this condition in terms of $R$:
for $\alpha>0$, $\beta<0$ the 
range of the scalar curvature is bounded
\[
 R <
-\left[2\beta \ W_{0}  \left(-\frac{\exp(\tfrac{1}{2}+\gamma^{-1})}{2\mu^2\beta} \right) \right]^{-1},
\]
where $W_0$ is the upper branch of the Lambert W function. If $|\gamma| \ll 1$, the exponential in the argument is small, so the upper limit is
\eq
f_R >0  \quad \Rightarrow \quad R \lesssim \mu^2e^{-\frac{2\alpha+\beta}{2\beta}}  = e^1R_*
\qquad\qquad (\alpha> 0, \ \beta<0)
\eqe
Thus, the condition ensuring the positivity of the scalar mass (\ref{m^2cond}) is sufficient to ensure that $G_{\rm eff}>0$. 
If we were to consider positive $\beta$, we need only recognise that since the function $f_{R}(R)$  
is decreasing as it crosses the axis at $f_{R}(R=0)=1$  the smallest value it can reach is $f_{R}(R=R_*) = 1-2\beta R_*$.
The condition can thus be expressed as
\eq
f_R >0  \quad \Rightarrow \quad R_*< \frac{1}{2\beta} \qquad\qquad (\alpha> 0, \ \beta>0)
\eqe
which, as noted above, is easily satisfied with the choice $\gamma \ll 1$.
For negative $\alpha$ we find\footnote{
Since the inverse function $R(\chi)$ is multivalued, for $\alpha<0$, $\beta>0$ there is a second valid
region: $R>  -\left[2\beta \ W_{0}  \left(-\exp(\tfrac{1}{2}+\gamma^{-1})/(2\mu^2\beta) \right) \right]^{-1} \simeq e^1R_*$. However, this corresponds to an extremely large value of the scalar curvature.
}

\eq
R < \left\lbrace \begin{array}{ll}  -\left[2\beta \ W_{0}  \left(-\frac{\exp(\tfrac{1}{2}+\gamma^{-1})}{2\mu^2\beta} \right) \right]^{-1} & (\alpha<0, \ \beta<0) \\
-\left[2\beta \ W_{-1}  \left(-\frac{\exp(\tfrac{1}{2}+\gamma^{-1})}{2\mu^2\beta} \right) \right]^{-1} & (\alpha<0, \ \beta>0)
\end{array} \right. ,
\eqe
where $W_{0}$ and $W_{-1}$ indicate the upper and lower branches of the Lambert W function
respectively.
Since for large $x$, $W_0(x)\sim \ln(x)$,  and for small $x$, $W_{-1}(x) \sim  \ln(-x) $, when $|\gamma|\ll 1$, we have
\eq\label{alphanegCond}
R\lesssim -\frac{1}{2\alpha},
\eqe
as in the $\beta=0$ case i.e. $f(R)=R+\alpha R^2$. For $\beta>0$ this is a stronger upper bound than that in (\ref{PostbetaNumCases}).
For $\beta<0$, $\gamma$ is positive and so (\ref{alphanegCond}) is weaker than (\ref{m^2cond}), which  already restricts $R$ to exponentially small values.
One difference between this and the $f(R)=R+\alpha R^2$ model
is that the negative $\alpha$ case is not ruled out by 
the $f_{RR}$ condition, so can be considered as a viable parameter choice, albeit for a restricted range of values of $R$.
These constraints are summarised in Table \ref{ConstraintTable}.

\begin{table*}[tb!]
\begin{tabular}{ | c | c || c | c |}
\hline
\multicolumn{2}{|c||}{Parameters}  & Unitarity & $m_{\chi}^2>0$ \\
\hline & & & \\ [-1em] \hline
\multirow{2}{*}{$\alpha > 0$} & $\beta > 0$ &$R_*<1/2\beta$ & $ R_* < R< 1/2\beta$  \\ \cline{2-4}
  & $\beta < 0$ &$R<e^1R_*$&  $R< R_*$ \\ \hline
\multirow{2}{*}{$\alpha < 0$} & $\beta > 0$ & $R<-1/2\alpha, \quad R\gtrsim e^1R_* $ &    
$1/2\beta < R< R_* $
\\ \cline{2-4}
 & $\beta < 0$ &$R<-1/2\alpha$&  $R< R_*$ \\ \hline
\end{tabular}
\caption{The  unitarity and positive-squared-mass constraints on the allowed curvature range for different values of the parameters $\alpha$ and $\beta$,
using $|\gamma|=|\beta/\alpha|\ll 1$ and $|\mu^2\alpha|\ll 1$. $R_*$ is defined in (\ref{Rstar}).}
\label{ConstraintTable}
\end{table*}

As with many $f(R)$ models in the literature, the potential $V(\chi)$ is multivalued, with branches at 
the points $\chi=\chi_*$ (see Fig. \ref{Cons_beta_plots}). As long as the conditions 
derived above are satisfied, the field will not reach these critical points. 
In the case of negative $\beta$ (with $\alpha>0$) this amounts to a (large) upper limit of the value of the spacetime 
curvature for which the model can be considered valid, which is far away from the stable solution at 
$R=0$ and for the small values of $|\gamma|$ considered here, significantly larger than the curvature
encountered in neutron stars. 
However, for positive $\beta$, the potential has no stable minimum when $\alpha>0$ and the branch point occurs at 
the lower limit of the range of validity, corresponding to a value of $R$ much smaller 
than the characteristic curvature of a neutron star.
 In a realistic scenario, 
 this could be remedied by the presence of a matter term $T\neq 0$, 
 which would give rise to a minimum in the effective potential.
Since the model in this paper is considered phenomenologically as an (ultraviolet) modification
to General Relativity that is relevant in the presence of dense nuclear matter, 
and in reality neutron stars are not completely isolated but instead occur in astrophysical situations 
with a non-zero stress-tensor, the instability may be avoided in practice. This notwithstanding, 
in the remainder of this paper we will consider only negative values of $\beta$.

The results of this subsection are presented in Table \ref{ConstraintTable}. In particular we note that for $\beta>0$,
the condition ensuring unitarity --- equivalent to $f_R>0$ for $f(R)$ theories --- is satisfied for a wide range of 
curvature values when $\alpha$ is positive, but is restricted to values less than $-1/2\alpha$ (as in the $f(R)=\alpha R^2$ case)
when $\alpha<0$. In the latter case, however, the condition for positive squared mass is significantly tighter, so this choice of
parameters would lead to instabilities for all but a tiny range of curvature values 
in the absence of matter. Despite this, in the numerical work in Sec. \ref{NS} we shall consider both positive and negative values of $\alpha$, so
as to compare with other works in the literature.

\subsection{Observational constraints}\label{SubSec:Observations}

We begin this subsection by considering the 
fifth force due to the extra scalar degree of freedom of the 
$f(R)$ theory. This fifth force can affect the effective 
gravitational constant $G_{\rm eff}$ and gravitational  
redshift at the surface of a neutron star $z_s$. 
By performing a conformal transformation 
\eq
 \tilde{g}_{\mu\nu} = F^2(\phi)g_{\mu\nu}\;, \label{conf.trans}	
\eqe
where
\eq
 F^2(\phi)\equiv f_R({R})=e^{-2Q\phi/M_{\rm pl}},  \label{F}
\eqe
the action (\ref{f(R)}) can be written in the Einstein frame 
\begin{align}
\tilde{\mathcal{S}}&=\int d^4x\sqrt{-\tilde{g}}\left\{\frac{M^2_{Pl}}{2}\tilde{R}-\frac{1}{2}\partial_{\mu}
\phi\partial_{\nu}\phi-V(\phi)\right\}
+\int d^4x \mathcal{L}_{M}(F^{-1}(\phi)g_{\mu\nu},\psi_{M})\;,\label{Einframe}\\
V(\phi)&=\frac{M^2_{Pl}}{2}\frac{f_R({R}){R}-f({R})}{f_R^2(R)}\;,\label{Vphi}
\end{align}
where
a tilde
indicates quantities in the Einstein frame, $\psi_M$ stands for the matter fields and for $f(R)$ theories
\eq
Q=-1/\sqrt{6}\;.
\eqe
Varying the action 
(\ref{Einframe}) \wrt the scalar field $\phi$ yields (in the spherical symmetric case) 
\begin{equation}
 \frac{d^2\phi}{d\tilde{r}^2}+\frac{2}{\tilde{r}}\frac{d\phi}{d\tilde{r}}-
 \frac{dV_{\rm eff}}{d\phi}=0\;,\label{phiequ}
\end{equation}
where $\tilde{r}=e^{-Q\phi/M_{\rm pl}} r$ and the effective potential $V_{\rm eff}$ is 
\begin{equation}
 V_{\rm eff}(\phi)=V(\phi)+\rho^{\ast}e^{Q\phi/M_{\rm pl}}\;,\label{chamVeff}
\end{equation}
with the conserved energy density in the Einstein frame $\rho^{\ast}=
e^{3Q\phi/M_{\rm pl}}\rho$. The effective potential in a medium with the density 
$\rho_i$ has a minimum at $\phi=\phi_i$ which is the solution of  
$dV_{\rm eff}/d\phi=0$ with the corresponding mass  $m_i^2\equiv V_{{\rm eff},\phi\phi}(\phi_i)$. 
The chameleon mechanism \cite{Khoury:2003aq,*Khoury:2003rn} can be described as follows. 
Inside the star ($\rho=\rho_{in}$), the chameleon field is almost frozen  at its minimum 
value $\phi_{\rm in}$, with corresponding mass $m_{\rm in}$ determined 
by the internal density $\rho_{\rm in}$. Then near the surface at $\tilde{r}_{1}<\tilde{r}_{s}$ (where 
$\tilde{r}_s$ is the star radius), the chameleon field
changes suddenly. Outside the star the scalar field 
is close to its minimum value $\phi_{\rm out}$, with corresponding mass $m_{\rm out}$
determined 
by the outside density $\rho_{\rm out}$. The  exact form of the chameleon field outside 
the star can be written as \cite{Khoury:2003aq,*Khoury:2003rn,Mota:2006fz}
\begin{equation}
 \phi(\tilde{r})\simeq\phi_{\rm out}-\frac{Q_{\rm eff}M_s}{4\pi M_{\rm pl}\tilde{r}}
 e^{-m_{\rm out}(\tilde{r}-\tilde{r_s})}\hspace{2cm} \tilde{r}>\tilde{r}_s\;,\label{cham.out}
\end{equation}
where $M_s$ is the total mass of the star and $\epsilon_{\rm th}$ is 
the thin shell parameter 
\begin{equation}
 \epsilon_{\rm th}=\frac{\phi_{\rm in}-\phi_{\rm out}}{6QM_{\rm pl}\Phi_{s}}\;,\label{eps}
\end{equation}
and the Newtonian potential at the surface of the star is $\Phi_{s}=\frac{GM_{c}}{\tilde{r}_{s}}$. 
The effective coupling constant $Q_{\rm eff}$ 
is defined as $Q_{\rm eff}=3Q\epsilon_{th}$ in the thin-shell regime ($\epsilon_{th}\ll1$)  and
$Q_{\rm eff}=Q$ in the thick-shell regime ($\epsilon_{th}\simeq\mathcal{O}(1)$).
The thin-shell parameter $\epsilon_{th}$, is an essential parameter of the chameleon mechanism.
This parameter determines if the modified theory satisfies the local constraints or not. For example, 
the post-Newtonian parameter $\gamma_{\rm PPN}$ is given by 
\begin{equation}
 \gamma_{\rm PPN}\simeq \frac{1-6Q^2\epsilon_{th}}{1+6Q^2\epsilon_{th}(1-\frac{\tilde{r}}{\tilde{r}_s})}\;,
\end{equation}
so that for $\epsilon_{th}\ll 1$, $\gamma_{\rm PPN}\simeq1$ as expected \cite{soti,Brax:2008hh}. 

\par For brevity, in the remainder of this section we drop the tilde on  
quantities in the Einstein frame. The
  force mediated by the chameleon field on a test body of mass $m$ at distance $r$ from a central body 
of mass $M_s$ and radius $r_s$ is\footnote{
The geodesic equation in the Jordan frame is: 
\begin{equation}
 \ddot{x}^{\mu}+\Gamma^{\mu}_{\alpha \nu}\dot{x}^{\alpha}\dot{x}^{\nu}=0\;, \nonumber
\end{equation}
and in the Einstein frame:
\begin{equation}
 \ddot{x}^{\mu}+\tilde{\Gamma}^{\mu}_{\alpha \nu}\dot{x}^{\alpha}\dot{x}^{\nu}=-\theta_{,\phi}\phi^{,\mu}-2\theta_{,\phi}\dot{x}^{\nu}
\dot{x}^{\mu}\phi_{,\nu} \;,\nonumber
\end{equation}
where $\theta\equiv \frac{Q}{M_{Pl}}\phi$. In the nonrelativistic limit the last 
term can be neglected and the chameleon force $\vec{F}_{ch}$ on a test particle is given by
\begin{equation}
\vec{F}_{ch}=-m\theta_{,\phi}\vec{\nabla}\phi\;, \nonumber
\end{equation}
} 
\begin{equation}
 |\vec{F}_{ch}|=m\frac{Q}{M_{Pl}}|\vec{\nabla}\phi|\;,\label{chamforce}
\end{equation}
where $\phi$ is given in (\ref{cham.out}). One can write for the total force (gravitational and chameleon)
\begin{equation}
 F_{tot}\equiv F_G+F_{\phi}=G_{\rm eff}\frac{mM_c}{r^2}\;,\label{niroo}
\end{equation}
where the effective gravitational coupling constant  is defined as:
\begin{eqnarray}
 G_{\rm eff}&\equiv&(1+\delta^2)G\;,\\ \label{sabethin}
 \delta^2&\simeq&2QQ_{\rm eff}\exp(-m_{out}({r}-{R}_s))\;, \label{delta}
\end{eqnarray}  
and G is the bare gravitational coupling constant

\begin{figure}[t]
\centering
\begin{tabular}{ccc}
\epsfig{file=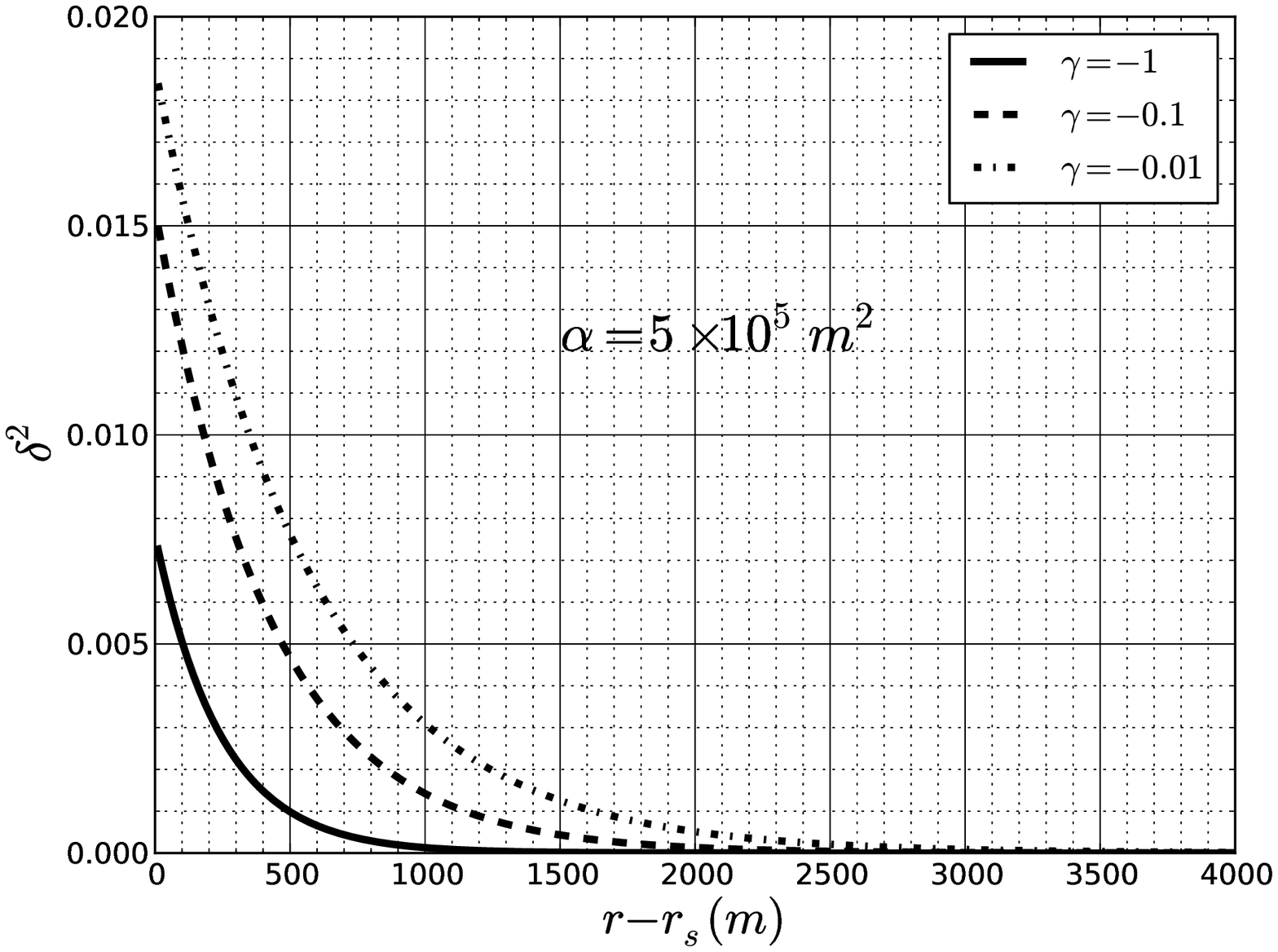,width=0.5\linewidth}& 
\epsfig{file=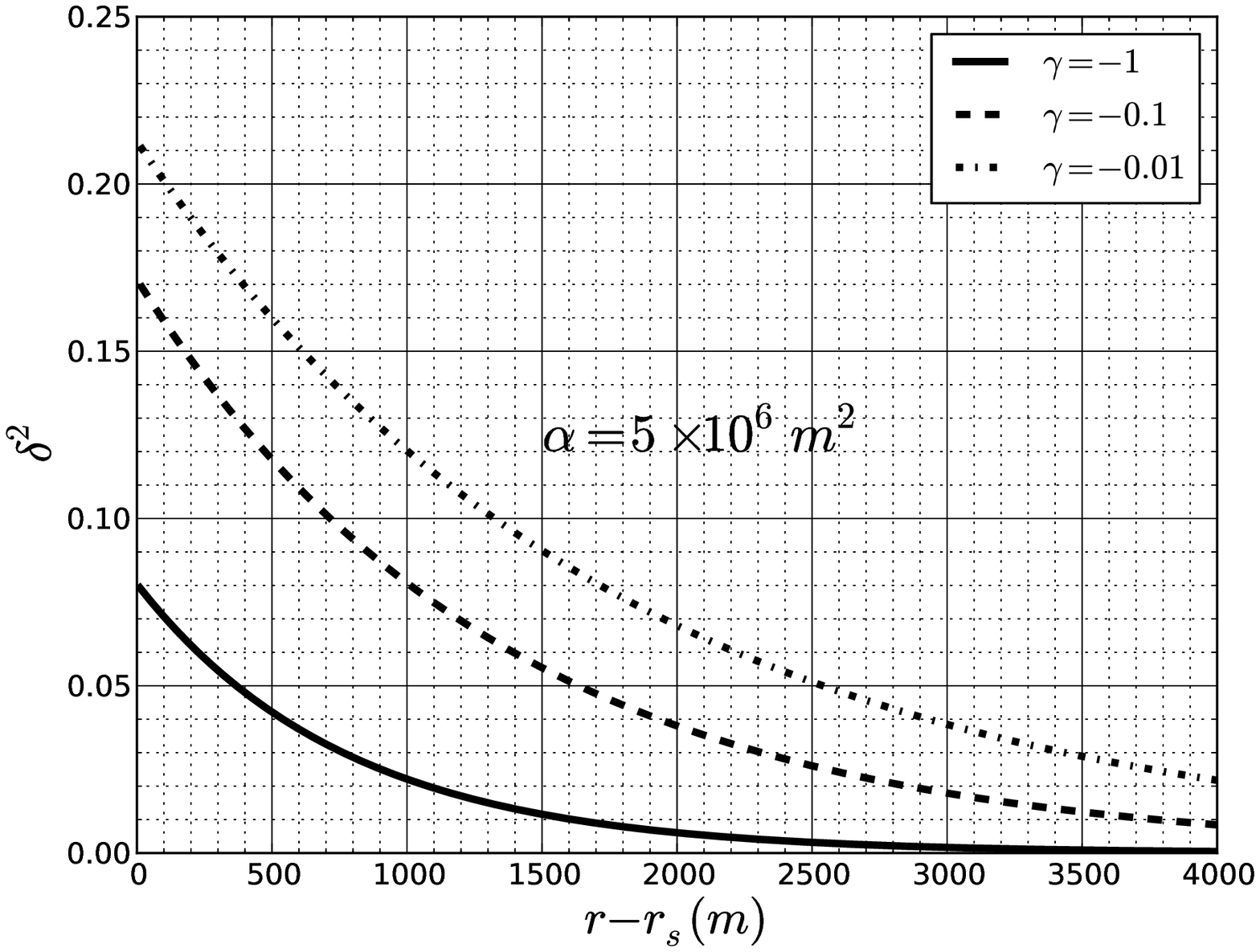,width=0.5\linewidth} \\
\end{tabular}
\caption{The  parameter $\delta^2\equiv G_{\rm eff}/G-1$ against 
the distance to the surface of a neutron star of radius $r_s=11{\rm km}$ and 
$M_s=2M_{\odot}$ 
in the $f(R)=R+\alpha R^2+\beta R^2\ln(R/\mu^2)$ 
gravity for different values of 
$\alpha$ and $\gamma\equiv\beta/\alpha$. 
}\label{Geff.fig}
\end{figure}

\par The parameter $\delta$ can be constrained with binary pulsar tests \cite{Esposito}.
For example, observations of the 
famous Hulse-Taylor binary pulsar PSR B1913+16 \cite{1913} give $|\delta|<0.04$.
The binary pulsars PSR J141-6545 \cite{1141} and PSR1534+12 \cite{1534} give $|\delta|<0.024$ and $|\delta|<0.075$ respectively.

The parameter $\delta^2$ for a  neutron star of mass $M=2M_{\odot}$ 
and radius $r_s=11{\rm km}$ for two values of parameter $\alpha$ and fixed \g$=\beta/\alpha$
is plotted in 
Fig. \ref{Geff.fig}. In this figure one can see that for the case with
 $\alpha=5\times10^5$,
  $\delta^2\lesssim 0.001$  for $r\gtrsim 1.2r_s$, so the 
 model easily satisfies the observational  constraints quoted above.
 For the larger value, $\alpha=5\times10^6$, $\delta^2$ takes larger values further from the surface
 of the star, however, since
binary pulsar tests are sensitive to the scale $r_{bs}\gg r_s$,
corresponding
of the order of the mean separation of the two stars,
any effect on the orbital motion of a binary system is completely negligible.\footnote{
One could also consider gravitational radiation from binary pulsars as
a potential discriminant between GR and modified gravity \cite{Damour:1998jk}.
It has been shown in \cite{radiation} that an application of $f(R)=R+\alpha R^2$ to the gravitational radiation of a 
hypothetical binary pulsar system requires 
that $\alpha<1.7\times10^{17}{\rm m^2}$ , under the assumption that 
the dipole power accounts for at most 1\% of the quadrupole power. 
However, as we shall see in the following section, consistent application of the perturbative
method means that we must restrict
$\alpha$ to values $\alpha\lesssim10^6 {\rm m^2}$. 
Thus, as far as our assumption that the logarithmic term constitutes only a
subdominant correction to the $R^2$ term holds true, the $f(R)$ model considered here
is not significantly constrained by measurements of the orbital period decay of double neutron stars.
}

\par However,
near the surface of the star, the deviation from GR is larger:
this deviation has observational effects on redshift of surface atomic lines  
that could in principle distinguish GR from modified theories of gravity
\cite{DeDeo:2003ju,*Psaltis:2007rv}.
The thermal spectrum of a neutron star will be detected by an observer 
at infinity with a gravitational redshift $z_s$ equal to
\begin{equation}
z_s\equiv \frac{\delta\lambda}{\lambda_{0}}=B(r)^{-1/2}-1\label{redshift}
\end{equation}
where $B(r)=1-2GM/r$ and $\lambda_{0}$ is 
the wavelength in the laboratory. Buchdahl's theorem \cite{buchlim} limits the 
value of $M/R$ for a spherical symmetric star in GR to $M/R<4/9$, 
so the maximum possible value of the of 
the redshift from the surface is $z_{s}\leq 2$.

\par
\begin{figure}[t]
\centering
\begin{tabular}{c}
\epsfig{file=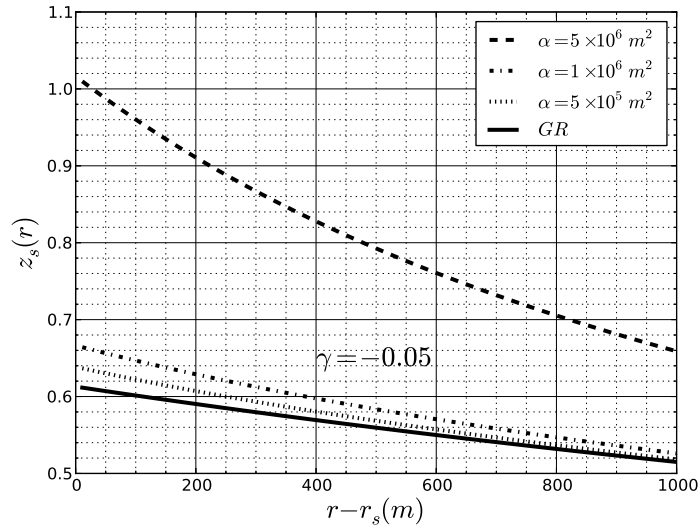,width=0.6\linewidth} 
\end{tabular}
\caption{The gravitational redshift parameter $z_s$ against 
the distance to the surface of a neutron star with radius $r_s=11{\rm km}$ and 
$M_s=2M_{\odot}$ in the 
$f(R)=R+\alpha R^2+\beta R^2\ln(R/\mu^2)$ model for different values of 
$\alpha$ and $\gamma\equiv\beta/\alpha=-0.05$.}\label{redshift.fig}
\end{figure}

 In Fig. \ref{redshift.fig} we have plotted $z_s$ as a function of $r$
in the immediate vicinity of the surface of a typical neutron star 
with mass $M_s=2M_{\odot}$ and radius $r_s=11{\rm km}$  for 
$\gamma=\beta/\alpha=-0.05$.
We can see that in this 
case of $\alpha=5\times10^{6}{\rm m^2}$, the deviation from 
GR is considerable, but for $\alpha=10^{6}{\rm m^2}$ 
and $\alpha=5\times10^{5}{\rm m^2}$, the gravitational redshift 
$z_s$  is close to the GR value 
$z^{GR}_s\simeq 0.51$. 
A large number of neutron stars exhibiting thermal emission have been observed by
X-ray satellites such as the Chandra X-ray Observatory, and XMM-Newton (see 
\cite{Ozel:2012wu} for a recent review)
and proposed missions such as ATHENA \cite{ATHENA} promise an increase
in the number and quality of the lines that can be used to analyse neutron star properties.
In principle then, for large $\alpha$ this deviation could be observed in 
lines originating close to the surface of the neutron star; in practice this would be 
dogged by uncertainties relating to the composition of the outer envelope of the neutron star, and
would require a careful treatment that is beyond the scope of this paper.

\section[Relativistic Stars]{The structure of relativistic stars}\label{NS}

As mentioned in the introduction, neutron stars probe the 
dense QCD phase diagram at low temperature and high 
baryon densities, where the baryon density in 
the stellar interior can  reach an order
of magnitude beyond the nuclear saturation density
 $\rho_{ns}=2.7\times 10^{17}\textrm{kg m}^{-3}$.
In such densities, matter can pass into a regime where the 
quark degrees of freedom are exited.
In this section we consider the internal structure of
relativistic stars within the framework of the phenomenological 
$f(R)$ model (\ref{f(R)})
and calculate the effect on 
the neutron star mass-radius (M-R) relation.

\subsection{Field Equations}

\par To obtain the field equations, we will use the method of perturbation constraints
adopted by Cooney et al. \cite{cooney}
for the study of neutron stars in $f(R)$ theory, and later used (in a slightly different form)
 by other authors 
\cite{Arapoglu,Orellana:2013gn,deli,Cheoun:2013tsa}.
This method is useful for investigating corrections to GR that give rise to
 field equations that would otherwise be almost unmanageable. 
The correction terms are 
treated as next to leading order terms in a larger expansion. 
To this end, 
the modified theory in Eq. (\ref{f(R)4}) is rewritten as
\begin{subequations}\label{f(R)3}
\begin{align}
 \mathcal{S}&=\frac{M_{Pl}^2}{2}\int d^4x\sqrt{-g}(R+\alpha h(R))
+\mathcal{S}_{m}\label{f(R)3.1}\;, \\
h(R)&=R^2+\gamma R^2\ln\frac{R}{\mu^2}\;,\label{f(R)3.2}
\end{align}
\end{subequations}
where we consider values such that $\gamma\equiv\beta/\alpha \ll 1$,
so the logarithmic term is a subdominant correction to the $R^2$ term.
The field equations arising from the action  (\ref{f(R)3}) are
\begin{eqnarray}
\label{EOM1}
R_{\mu \nu}-\frac{1}{2}g_{\mu \nu}R + 
\alpha \left[h_R R_{\mu \nu} - \frac{1}{2}g_{\mu \nu}h - \left(\nabla_{\mu}\nabla_{\nu} 
-g_{\mu \nu}\square \right)h_R \right]= 8\pi G T^{m}_{\mu \nu}\;,
\end{eqnarray}
where $h_R \equiv \delta h / \delta R$ and 
$T^{m}_{\mu\nu}\equiv -2/\sqrt{-g}\partial\mathcal{S}_{m}/\partial g^{\mu\nu} $. 
Taking the trace of Eq. (\ref{EOM1}) 
\begin{equation}
\label{TR}
R - \alpha \left[h_R R - 2h +3\square h_R \right]= -8\pi GT\;,
\end{equation}
and substituting $R$ from Eq. (\ref{TR}) in to Eq. (\ref{EOM1}) gives
\begin{eqnarray}
\label{EOM2}
R_{\mu\nu}+ \alpha \left[h_R R_{\mu \nu} - \frac{1}{2}g_{\mu \nu}\left(h_R R-h\right) -
 \left(\nabla_{\mu}\nabla_{\nu} +\frac{1}{2}g_{\mu \nu}\square \right)h_R \right]
=8\pi G\left(  T^{m}_{\mu \nu} -\frac{1}{2}g_{\mu \nu}T^{m}\right)\;.
\end{eqnarray}
We shall consider the perturbative expansion in the dimensionless constant 
\eq
c_R =\alpha \mu^2
\eqe
 (recall from (\ref{muDef}) that $\mu^2$ is of the order of the curvature of a typical neutron star).
At zeroth order in $c_R$, the equations are ordinary GR equations with 
$g_{\mu\nu}^{(0)}$ solutions; 
in the perturbative approach we expand the quantities in the metric
and stress-energy tensor up to first order in $c_R$ i.e.
\begin{equation}
\label{expansion}
g_{\mu \nu } = g^{(0)}_{\mu\nu}+c_R g^{(1)}_{\mu\nu}\;.
\end{equation}
Considering the line element 
\begin{equation}
\label{METRIC}
ds^2 = -B(r)dt^2 + A(r)dr^2+r^2\left( d\theta^2 + \sin^2\theta d\phi^2\right) ,
\end{equation}
and assuming a perfect fluid inside the 
star ($T^{m\mu}_{\nu}=\textmd{diag}[-\rho,P,P,P]$) the field equations  (\ref{EOM2})   can be written 
\begin{subequations}\label{Rii}
 \begin{align}
  \frac{R_{00}}{B}&+\alpha\left[h_{R}\frac{R_{00}}{B}+\frac{1}{2}(h_RR-h)+\frac{1}{2A}(h''_R
+(\frac{3B'}{2B}-\frac{A'}{2A}+\frac{2}{r})h'_R)\right]=4\pi G(\rho+3P)\label{R00}\;,\\
\frac{R_{11}}{A}&+\alpha\left[h_{R}\frac{R_{11}}{B}-\frac{1}{2}(h_RR-h)-\frac{1}{2A}(3h''_R
+(\frac{B'}{2B}-\frac{3A'}{2A}+\frac{2}{r})h'_R)\right]=4\pi G(\rho-P)\label{R11}\;,\\
\frac{R_{22}}{r^2}&+\alpha\left[h_{R}\frac{R_{22}}{B}-\frac{1}{2}(h_RR-h)-\frac{1}{2A}(h''_R
+(\frac{B'}{2B}-\frac{A'}{2A}+\frac{4}{r})h'_R)\right]=4\pi G(\rho-P)\;,\label{R22}
 \end{align}
\end{subequations}
where a prime indicates differentiation \wrt r. To first order in $c_R$ the pressure 
and the energy density are $P=P^{(0)}+c_R P^{(1)}$ 
and $\rho=\rho^{(0)}+c_R \rho^{(1)}$ respectively.

\subsection[Modified Tolmann-Opp...]{Modified Tolmann-Oppenheimer-Volkov equations}
\par In astrophysics, the Tolman-Oppenheimer-Volkoff (TOV) equations constrain the structure 
of a spherically symmetric body of isotropic 
material that is in static gravitational equilibrium \cite{TOV}. 
Before considering an ansatz for the solutions inside the star and 
obtaining the modified Tolmann-Oppenheimer-Volkov equations (MTOV), something should 
be said about the exterior solutions. As the modified theory in Eq. (\ref{f(R)3}) 
is considered for high curvature regimes in presence of matter, we assume that, outside of the star, 
the solutions can be approximately explained by the Schwarzschild solution 
\begin{equation}
 A_{out}(r)=B_{out}(r)^{-1}=\left(1-\frac{2GM_{tot}}{r}\right)^{-1}\;,\label{extsol}
\end{equation}
where for a few radii far from the star,  $M_{tot}$ receives no corrections
due to the modified theory. However for distances close to the surface of the star, a good approximation 
should include the \a corrections. 

\par The ansatz for the interior solutions is then 
\begin{equation}
\label{ADEF}
A(r)\equiv\left(1-\frac{2GM(r)}{r} \right)^{-1}\; ,
\end{equation}
where $M(r)$ contains corrections to the first order in $\alpha$ arising from the form of $h(R)$.
Using Eqs. (\ref{Rii}) and the geometrical relation
\begin{equation}
\label{MDEF}
\frac{R_{00}}{2B}+\frac{R_{11}}{2A}+\frac{R_{22}}{r^2} = \frac{2M'G}{r^2}\;,
\end{equation}
the first MTOV equation is found to be
\begin{equation}
 \frac{dM}{dr}=4\pi\rho r^2-\alpha r^2\left(4\pi\rho h_R-\frac{1}{4G}(h_RR-h)
-\frac{1}{2AG}\left((\frac{2}{r}-\frac{A'}{2A})h'_R+h''_R\right)\right)\; .\label{MTOV1}
\end{equation}
The second MTOV equation is derived by using Eq. (\ref{R22}), 
the conservation equation $\nabla_{\mu}T^{m\mu}_{\nu}=0$
\begin{equation}
\label{DT}
\frac{B'}{B} = -\frac{2P'}{\rho+P}\;,
\end{equation}
and the relation
\begin{equation}
\label{R22.2}
\frac{R_{22}}{r^2} = \frac{G}{r^2}\left[\frac{dM}{dr} + \frac{M}{r}-\frac{r}{A}(\frac{B'}{B})\right].
\end{equation}
This gives
\begin{equation}
\label{MTOV2}
 \frac{dP}{dr}=-\frac{A}{r^2}(\rho+P)\left[MG+4\pi Gr^3P-\alpha r^3\left(\frac{1}{4}(h_RR-h)
+\frac{1}{2A}(\frac{2}{r}+\frac{B'}{2B})h'_R+4\pi G Ph_R\right)\right]\; .
\end{equation}

\subsection{Neutron stars}
\par The structure of neutron stars has been
previously studied in \f models of the form $f(R)\sim R+\alpha R^2$ \cite{cooney,Arapoglu,Orellana:2013gn} and the 
Starobinsky model \cite{babi} as well as in models incorporating $R^{\mu\nu}R_{\mu\nu}$ terms
\cite{deli,Santos:2011ye} and the gravitational aether theory \cite{Kamiab:2011am}.  
The
modification to GR manifests itself in observable features such as the mass-radius (M-R) relation of 
neutron stars. 
To solve Eqs. (\ref{MTOV1}) and  (\ref{MTOV2}) a third equation is needed
to relate the
 matter density $\rho$ and the pressure $P$ i.e. 
the equation of state (EoS) of the neutron star. 
The EoS contains information about the behavior of the matter 
inside the star. As the properties of matter at high densities 
are not well known, there are 
different types of equation of state that give rise to different M-R relationships
\cite{Lattimer:2000nx,Steiner:2010fz}.
Here, we consider two types of EoS: the simpler polytropic EoS and a more realistic SLy EoS \cite{Haensel}.

\subsubsection{Polytopic EoS}

In this case we consider a simplified polytropic \eos

\begin{equation}
\zeta=2\xi+5.0 \;,\label{Polyeos}
\end{equation}
where 
\begin{equation}\label{xizetaDefs}
\xi=\log(\rho/\textrm{g cm}^{-3}),\qquad\qquad
\zeta = \log(P/\textrm{dyn}\;\textrm{cm}^{-2})\;.
\end{equation}
The MTOV 
 equations (\ref{MTOV1}) and (\ref{MTOV2}), together with (\ref{Polyeos}), were then solved numerically, 
 using a Fehlberg fourth-fifth order Runge-Kutta method to integrate from the center 
 of star to the surface. We define the surface of the star as the point
 where the density drops to a value of order $10^9{\rm  kg/m^3}$. In the \f model in hand, 
 $h_{RR}$ includes the $\ln(R/\mu^2)$ term, which is not well defined at $R=0$. Thus, we restrict the 
 calculation to the $R>0$ domain. 
 
\begin{figure}[th!]
\centering
\begin{tabular}{cc}
\epsfig{file=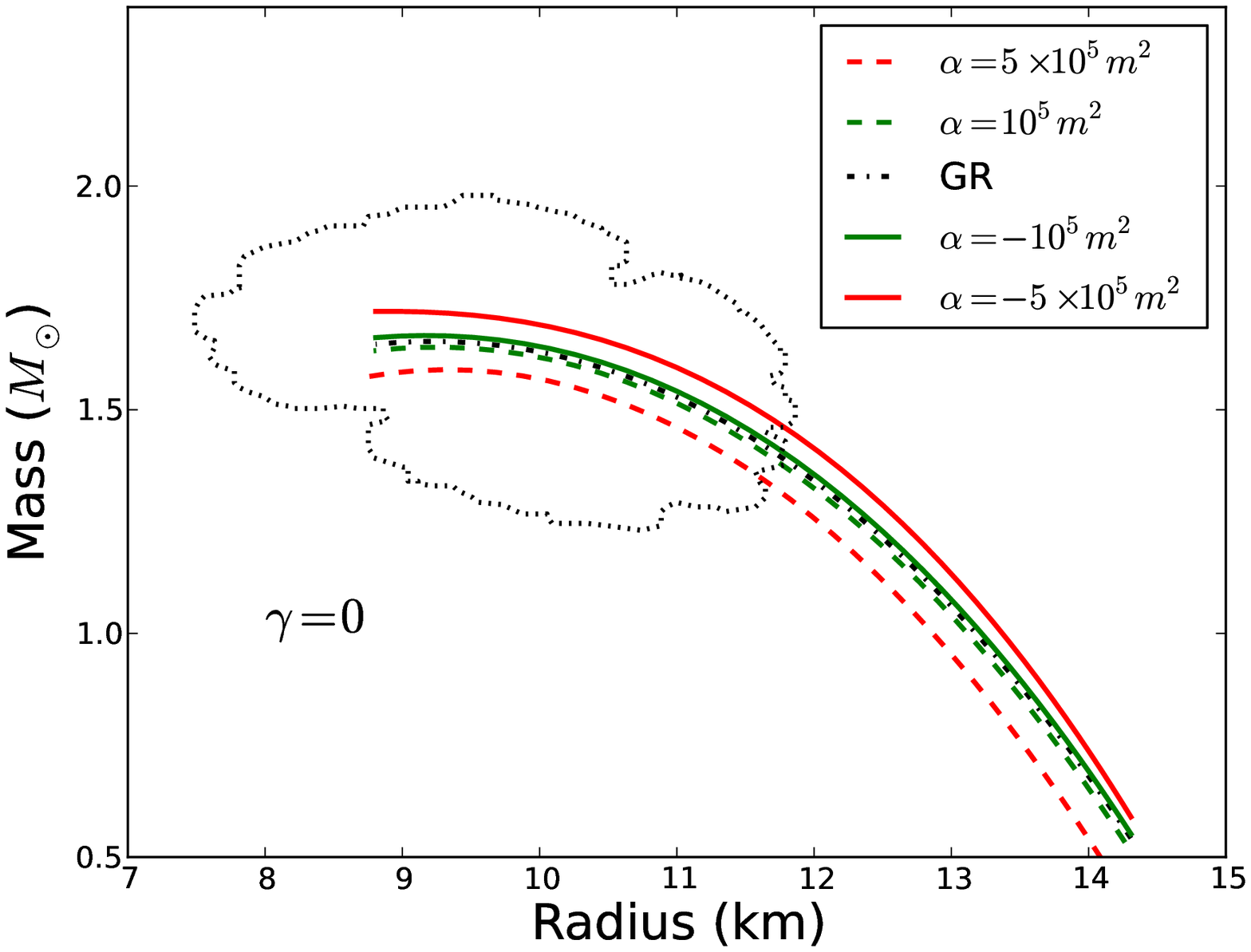,width=0.5\linewidth,clip=} & \epsfig{file=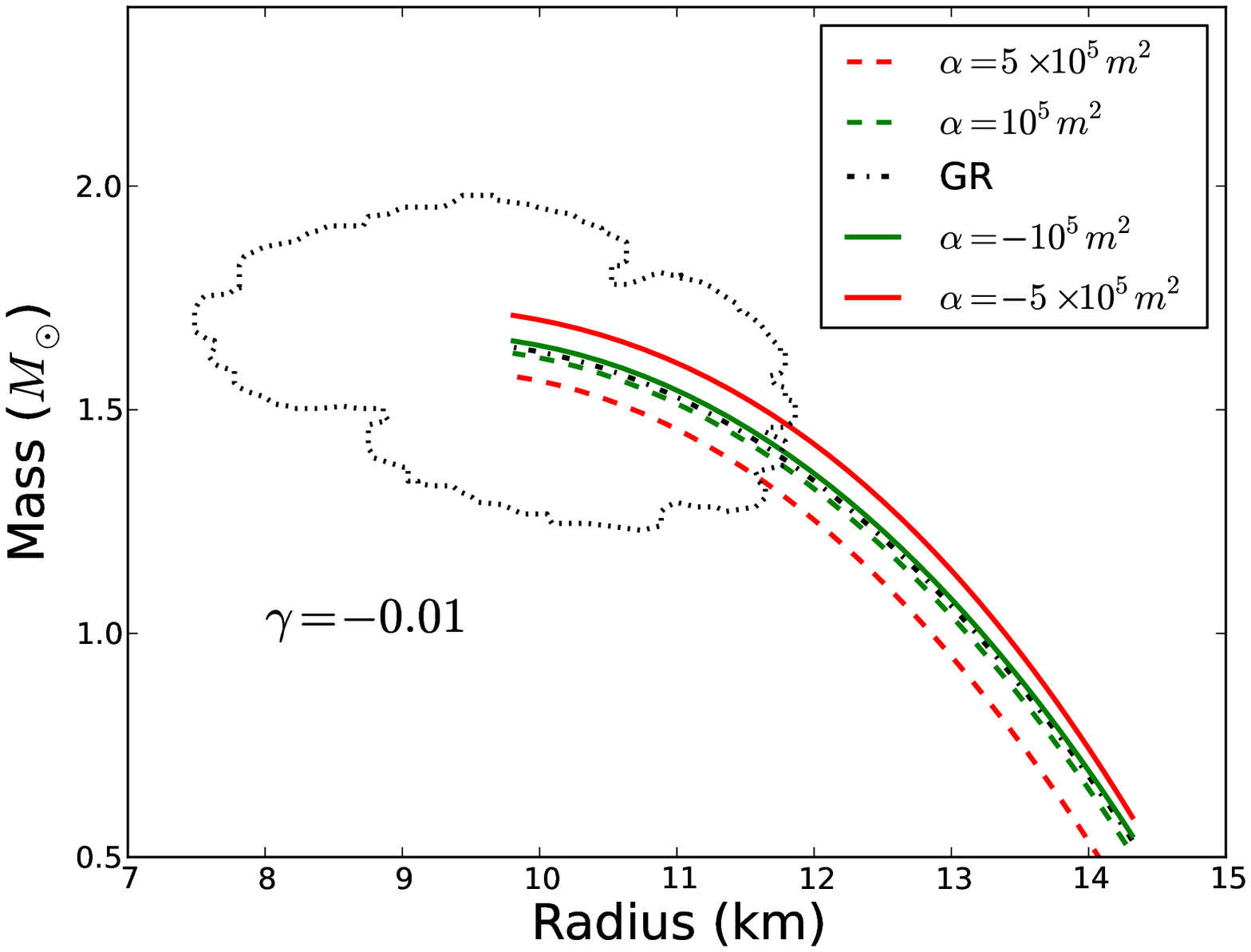,width=0.5\linewidth,clip=}\\
\epsfig{file=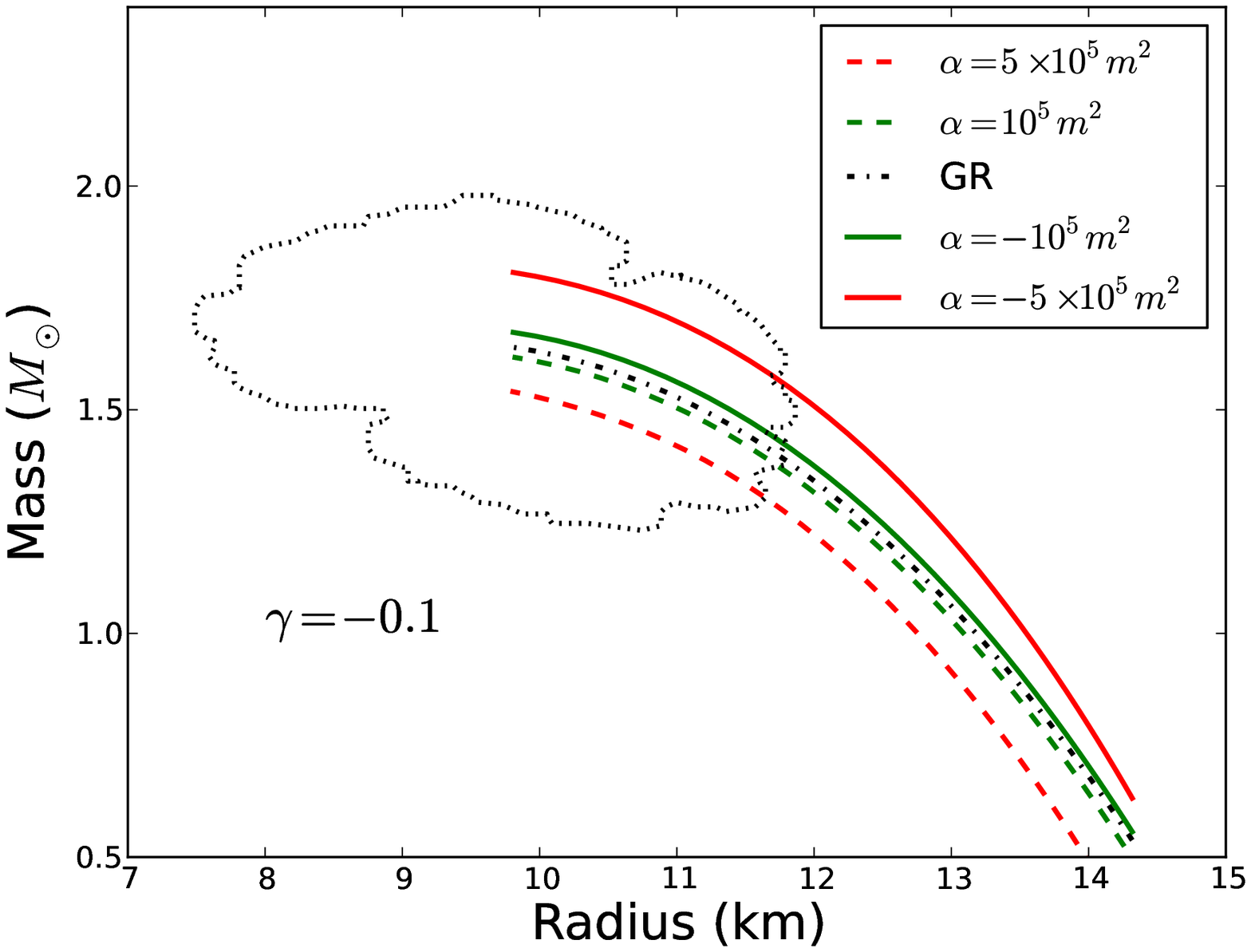,width=0.5\linewidth,clip=} & \epsfig{file=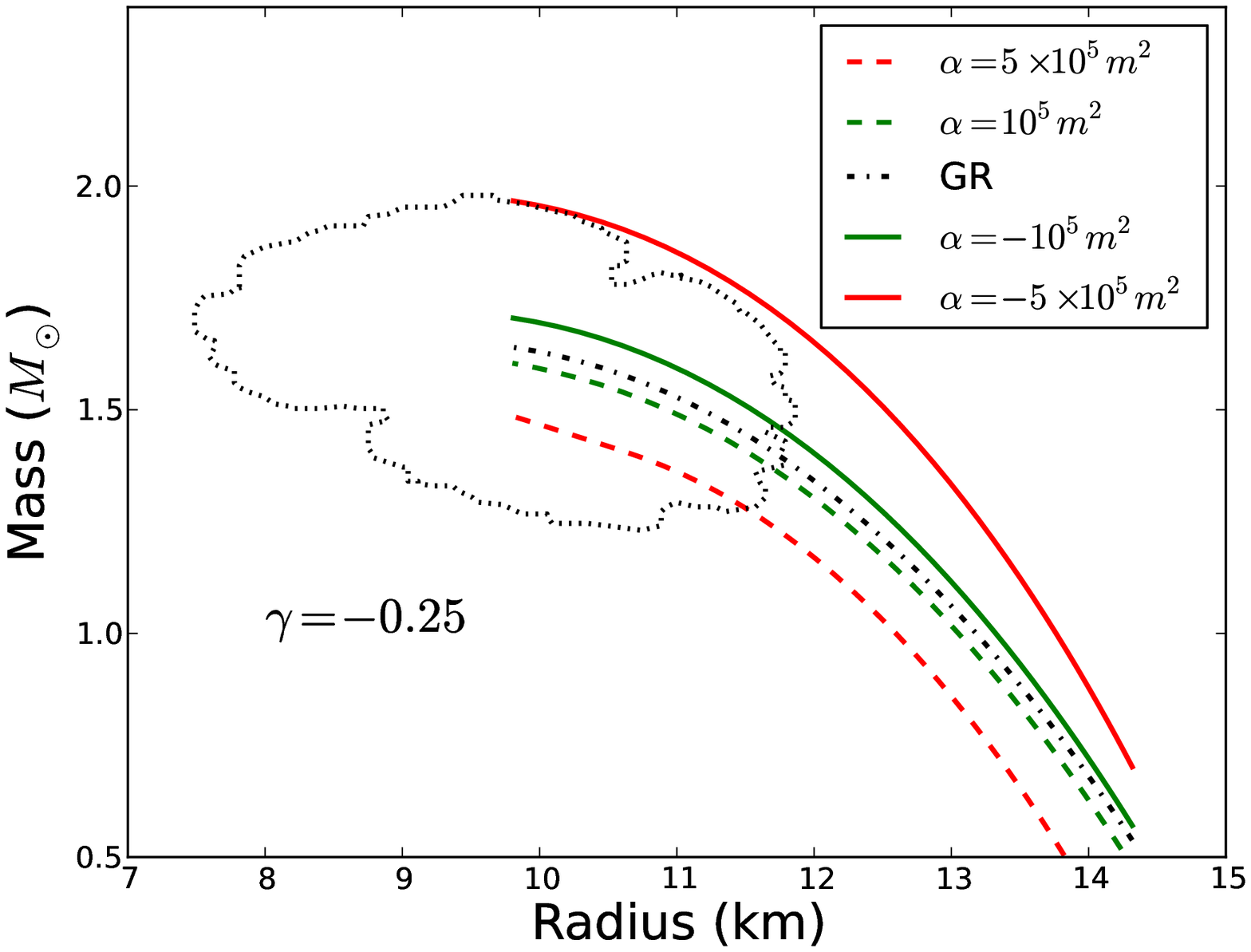,width=0.5\linewidth,clip=}\\
\end{tabular}
\caption{(Color online.) The mass-radius (M-R) diagram for neutron stars in GR ($\alpha=\beta=0$)
and $f(R)=R+\alpha R^2+\beta R^2\ln R/\mu^2$ 
 using a simplified polytropic \eos (\ref{Polyeos}).
Here $\gamma\equiv\beta/\alpha$ and the range of the  matter density at 
the center of the  star is varied from $\rho_{ns}$ to the point where the 
Ricci scalar goes to zero for the $\gamma\neq 0$ cases. $\rho_{ns}=2.7\times 10^{17}\textmd{kg\;m}^{-3}$ is the nuclear saturation density. 
The dotted contour gives the $2\sigma$ constraints derived from
observations of three neutron stars 
 reported in \cite{ozel}. The presence of the logarithmic term ($\gamma\neq0$) can be seen to cause  
 larger deviations from the GR case compared to the R-squared model ($\gamma=0$).}\label{PolyMRdiag}
\end{figure}

 The density at the center of star is increased
from $\rho_{ns}$ ($\rho_{ns}=2.7\times 10^{17}\textmd{kg\;m}^{-3}$ is the nuclear saturation density)
 until the point where the 
Ricci scalar goes to zero.  
The numerical results for this case 
are shown in Fig. \ref{PolyMRdiag}. 
In this case the deviation from GR can clearly be seen to increase for larger values of 
$\gamma$. For this type of \eos it can also  be seen that the deviation from GR 
 becomes more asymmetric for negative and 
positive values of \a as $\gamma$ increases, and positive (negative) values of $\alpha$ give rise to lower (higher) mass stars for a given radius.

\subsubsection{SLy EoS}

\begin{figure}[t]
\centering
\begin{tabular}{cc}
\epsfig{file=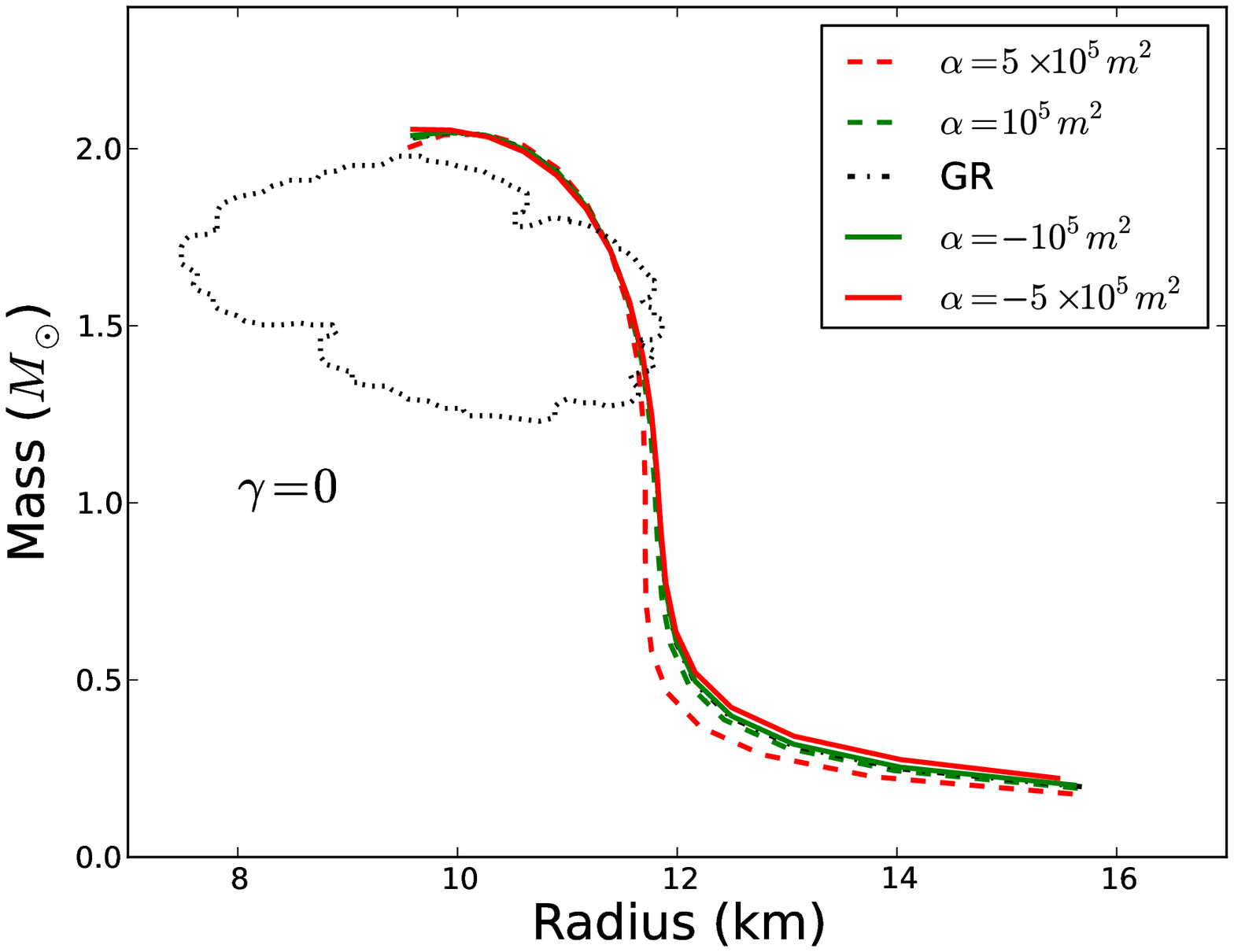,width=0.5\linewidth,clip=} & \epsfig{file=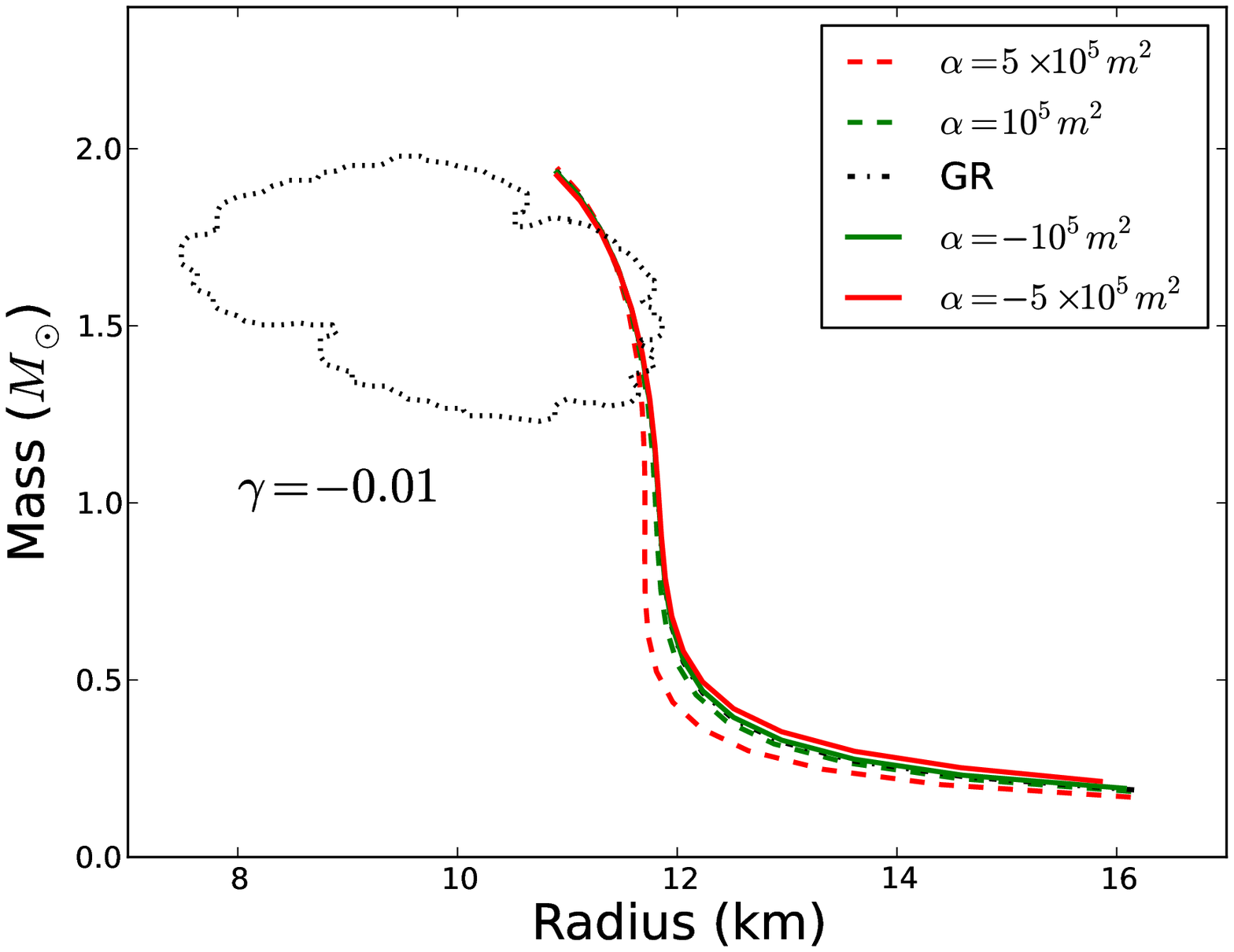,width=0.5\linewidth,clip=}\\
\epsfig{file=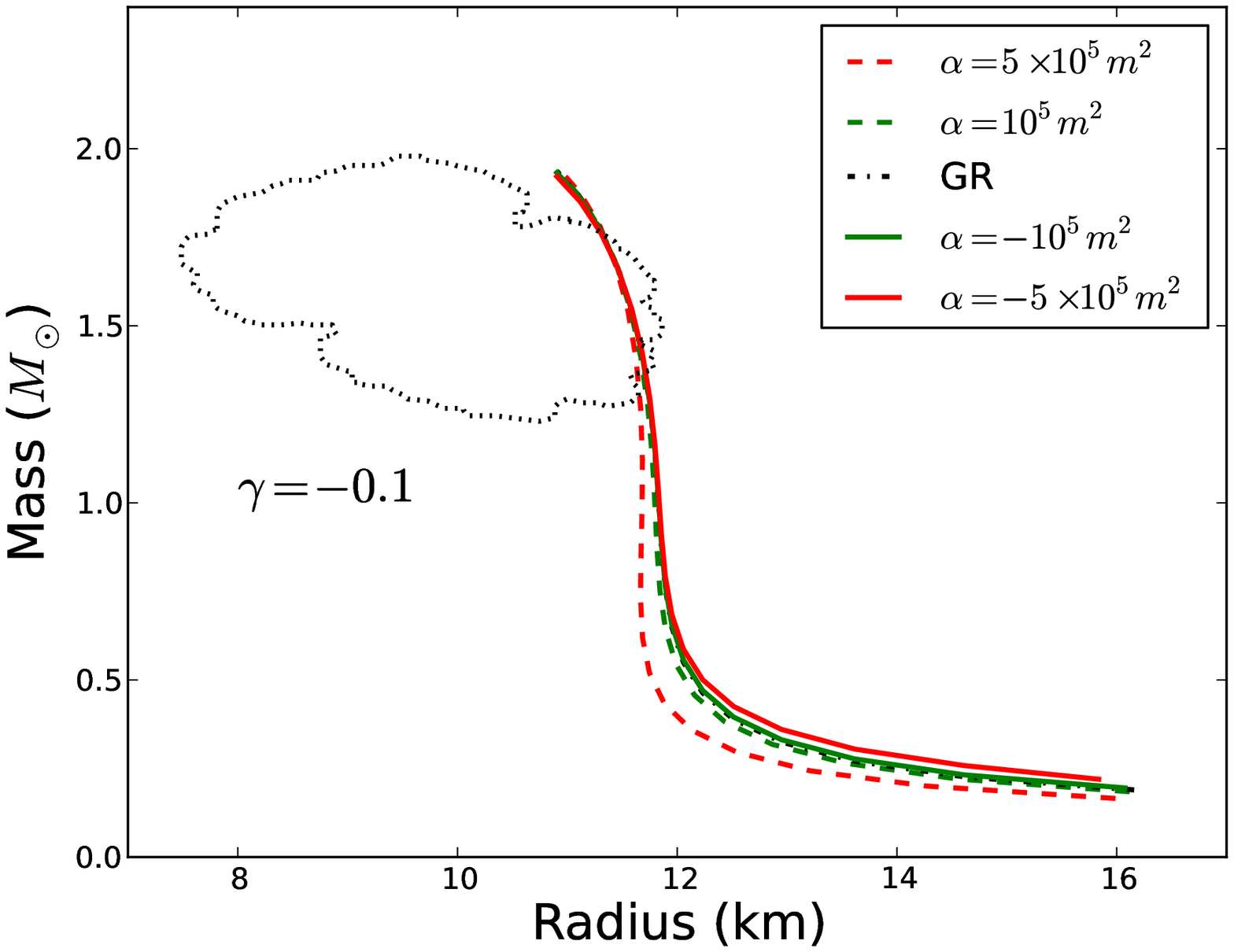,width=0.5\linewidth,clip=} & \epsfig{file=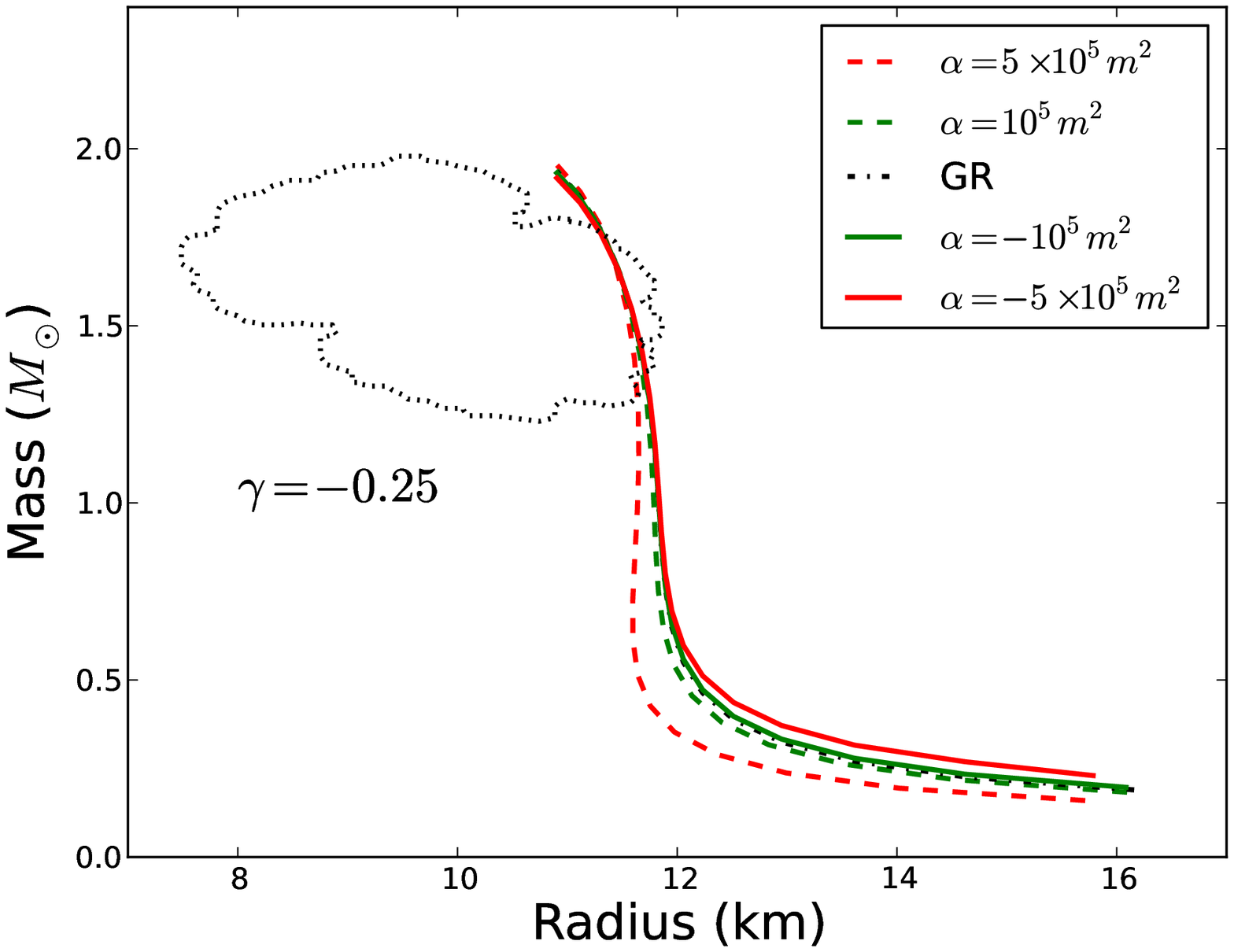,width=0.5\linewidth,clip=}\\
\end{tabular}
\caption{(Color online.) The mass-radius (M-R) diagram for neutron stars in GR ($\alpha=\beta=0$)
and $f(R)=R+\alpha R^2+\beta R^2\ln R/\mu^2$ 
 using the realistic SLy equation of state (\ref{SLyeos}).
Here $\gamma\equiv\beta/\alpha$ and the range of the  matter density at 
the center of the  star changes from $\rho_{ns}$ to the point where the 
Ricci scalar goes to zero for the $\gamma\neq0$ cases. $\rho_{ns}=2.7\times 10^{17}\textmd{kg\;m}^{-3}$ is the nuclear saturation density. 
The dotted contour gives the $2\sigma$ constraints derived from
observations of three neutron stars 
 reported in \cite{ozel}. For larger values of $\gamma$, the presence of the logarithmic term can be seen to cause  
 larger deviations from the GR case compared to the R-squared model ($\gamma=0$).
 The deviation from the GR case is most prominent where the central density (and thus the pressure) takes intermediate values such that
 $R$ is large.}\label{SLyMRdiag}
\end{figure}

The SLy \eos models the behavior of nuclear matter at high densities . 
An explicit analytic representation is
\begin{eqnarray}
  \zeta&=&\frac{a_1+a_2\xi+a_3\xi^3}{1+a_4\,\xi}\,f_0(a_5(\xi-a_6))+ (a_7+a_8\xi)\,f_0(a_9(a_{10}-\xi))
\nonumber\\&&
     + (a_{11}+a_{12}\xi)\,f_0(a_{13}(a_{14}-\xi)) + (a_{15}+a_{16}\xi)\,f_0(a_{17}(a_{18}-\xi)) \;.
\label{SLyeos}
\end{eqnarray} 
where $\xi$ and $\zeta$ are defined as in (\ref{xizetaDefs}) and
\begin{equation}
f_0(x) = \frac{1}{\mathrm{e}^x+1}.
\end{equation}
The coefficients $a_i$ are listed in \cite{Haensel}. 
  The results are shown in Fig. \ref{SLyMRdiag}. Here again the density at the center of star changes from $\rho_{ns}$ to the point where the 
Ricci scalar goes to zero. 
As the SLy \eos is stiff and  $R\propto(\rho-3P)$, when $\gamma \ne 0$ we do not obtain stars with a radius smaller than $r_s\sim 11{\rm km}$, compared to $r_s<10{\rm km}$
for the R-squared model (left-top panel).
 The deviation from the GR case is most prominent where the central density (and thus the pressure) takes intermediate values such that
 $R$ is large. At this point, which corresponds to extremely low-mass stars, an asymmetric deviation from GR that increases in magnitude with $|\gamma|$ can be seen, as with the polytropic 
 equation of state.
 However, here it is the solutions corresponding to positive $\alpha$ that exhibit the greatest deviation from GR. 
 
 As in the $f(R)=R+\alpha R^2$ model \cite{Arapoglu,Orellana:2013gn}
there is an inversion of the modified gravity effect near the  central density
$\rho\simeq 5\rho_{ns}$ for the SLy equation of state.
 This point corresponds to stars with a mass $\sim 2M_\odot$; since this 
 is close to the point where $R=0$ (beyond which the logarithmic model is
 not valid) there is little deviation from the GR case for stars with astrophysical masses for this equation
 of state. 
 If one were to use a softer equation of state (which permits a larger range of central densities)
 one would expect larger deviations from the GR case after this inversion point.

\subsection{Quark stars}\label{Sec:QuarkStars}

The concept of a star made of strange quark matter was first 
suggested by Itoh \cite{1970PThPh..44..291I} and later expanded upon by Witten \cite{witten}. 
The unusual physical properties, such as the absence of
a minimum mass and a
finite density but zero pressure at their surface were later
studied by Alcock et al. \cite{Alcock:1986hz, Alcock:1988re}. 
In this model it is assumed 
that the star is made mostly of u, d, s quarks together with 
electrons, which give total charge neutrality.
The interior of the star is made up of deconfined quarks that form a colour superconductor,
leading to a softer equation of state
with possible observable effects on the minimum mass, radii, 
cooling behaviour and other observables \cite{Weber:2006iw,*Lattimer:2000nx,*Ozel:2006bv}. 
In this subsection we  investigate the effect 
of the modified gravity on the structure of this type of self-bound star.

\par
\begin{figure}[t]
\centering
\begin{tabular}{cc}
\epsfig{file=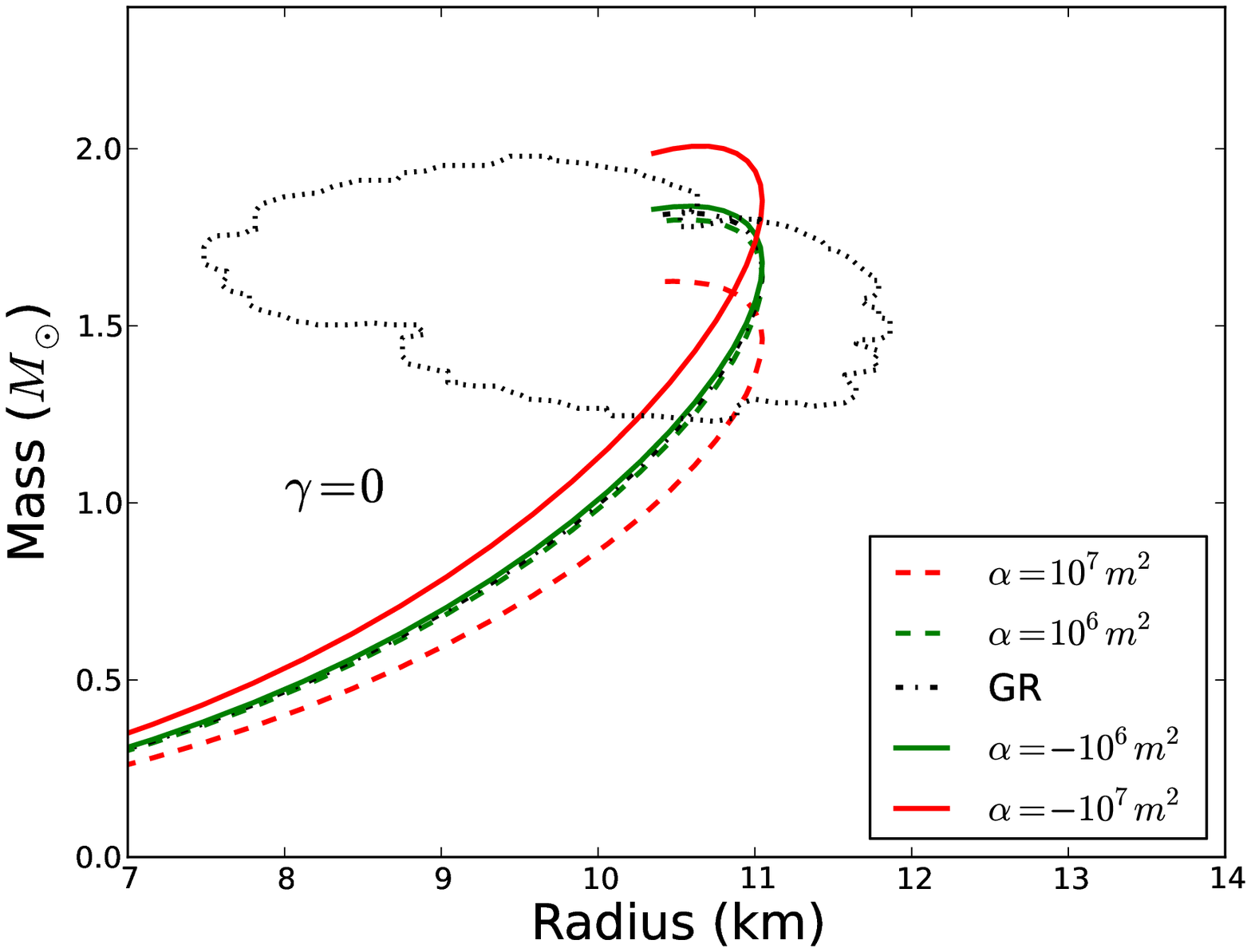,width=0.5\linewidth,clip=} & \epsfig{file=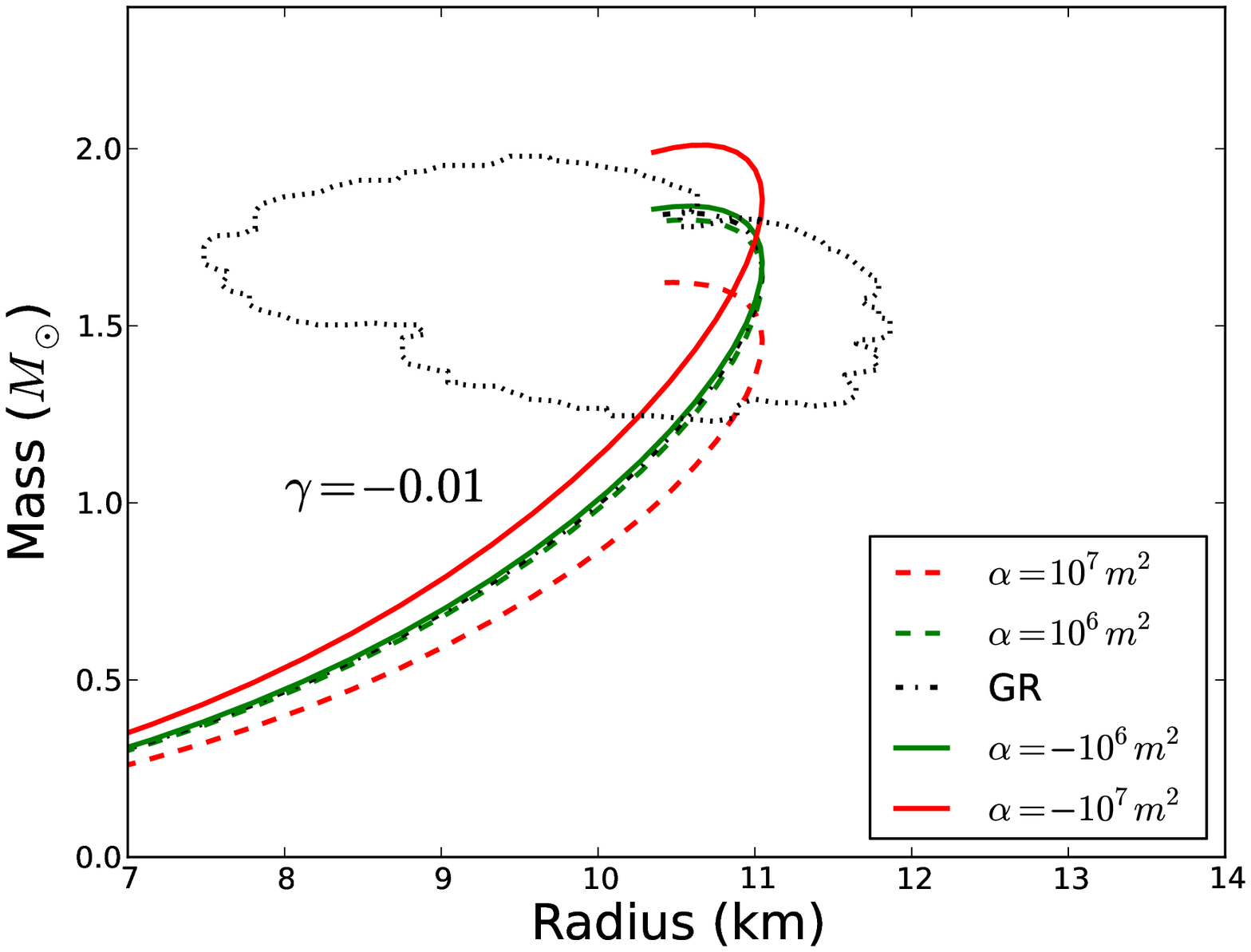,width=0.5\linewidth,clip=}\\
\epsfig{file=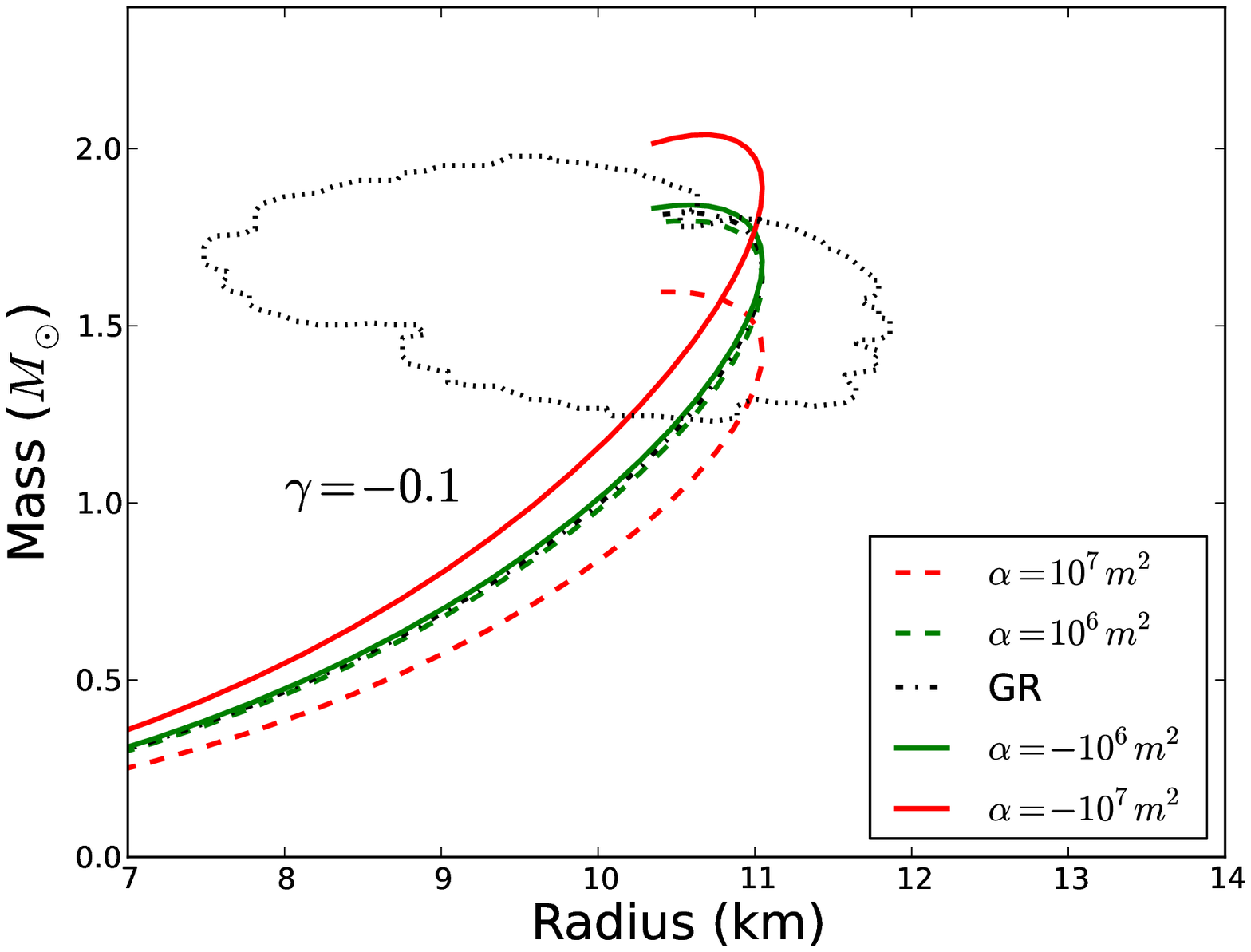,width=0.5\linewidth,clip=} & \epsfig{file=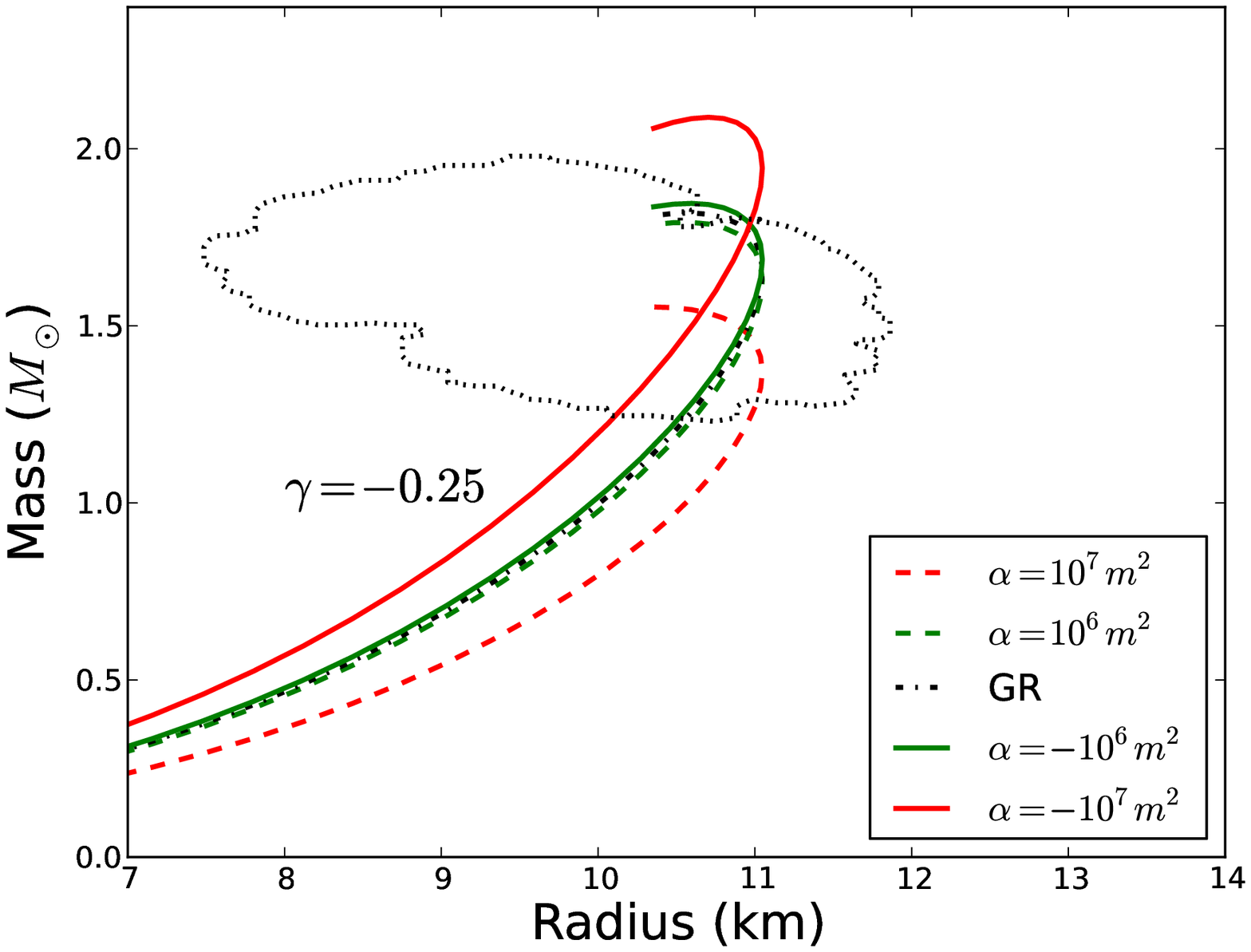,width=0.5\linewidth,clip=}\\
\end{tabular}
\caption{(Color online.) The mass-radius (M-R) diagram for the quark star case in GR and $f(R)=R+\alpha R^2+\beta R^2\ln R/\mu^2$ 
 using a linear \eos (\ref{QSEoS}) with $a=0.28$ and $\mathcal{B}=1$.
Here $\gamma\equiv\beta/\alpha$ and the range of the  matter density at 
the center of the  star changes from $1.54\rho_{ns}$ to $9.3\rho_{ns}$,
where $\rho_{ns}=2.7\times 10^{17}\textmd{kg\;m}^{-3}$ is the nuclear saturation density. 
The dotted contour gives the $2\sigma$ constraints derived from
observations of three neutron stars 
 reported in \cite{ozel}.}\label{QSMRdiag}
\end{figure}

 The equation of state 
of strange matter made up of u,d, s quarks can 
be considered  in the framework 
of the MIT bag model. In this model,
a linear approximation is assumed as  \cite{lrr-2003-3}
\begin{equation}
 P\simeq a(\rho-\rho_0)\;,\label{QSEoS}
\end{equation}
where $\rho_0$ is the density of the strange matter at zero pressure. 
The MIT bag model describing the strange quark matter involves three parameters, viz.
The bag constant $\mathcal{B}=\rho_0/4$, the strange quark mass $m_s$ and the 
QCD coupling constant $\alpha_c$. If we neglect the strange quark mass, then $a=1/3$. For 
$m_s=250$ MeV we have $a=0.28$. In units of 
$\mathcal{B}_{60}=\mathcal{B}/(60{\rm Mev\;fm^{-3}})$, the constant $\mathcal{B}$ is 
restricted to $0.98<\mathcal{B}<1.52$ \cite{lrr-2003-3}. The M-R diagram for a quark star with $a=0.28$ and $\mathcal{B}=1$ is 
shown in Fig. \ref{QSMRdiag}. From this figure it is clear that the masses of 
quark stars with negative values of $\alpha$ are always enhanced with respect to GR
and the masses of quark stars with positive values of $\alpha$ are diminished 
relative to GR, irrespective of the value of $\gamma$. Compared to the SLy and polytropic equations of state, 
 larger values of $\alpha$ [i.e. $\alpha= \mathcal{O}(10^7 m^2)$] can give rise to stars with masses and radii in the ranges allowed by the observational constraints. 
As in the previous subsection, it can be seen that the deviation is larger for larger
values of $|\gamma|$. In the case of the quark star, however, the \eos is  less stiff so there is more 
deviation in the mass-radius diagram with respect to GR.

\subsection{The perturbative regime}

 In all considered cases, it is important to stay in the perturbative regime, 
 so that the first order corrections to the metric in (\ref{expansion})
 are small.
 This can be measured quantitatively with 
\begin{equation}
 |\Delta| = \left|\frac{A_{MG}(r)}{A_{GR}(r)}-1\right|\;,\label{Polyperteq}
\end{equation}
where 
$A(r)$ is the rr component of the metric defined in Eq. (\ref{ADEF}) and the subscripts MG and GR refer to the modified gravity and 
General Relativity cases respectively.

\begin{figure}[t]
\centering
\begin{tabular}{ccc}
\epsfig{file=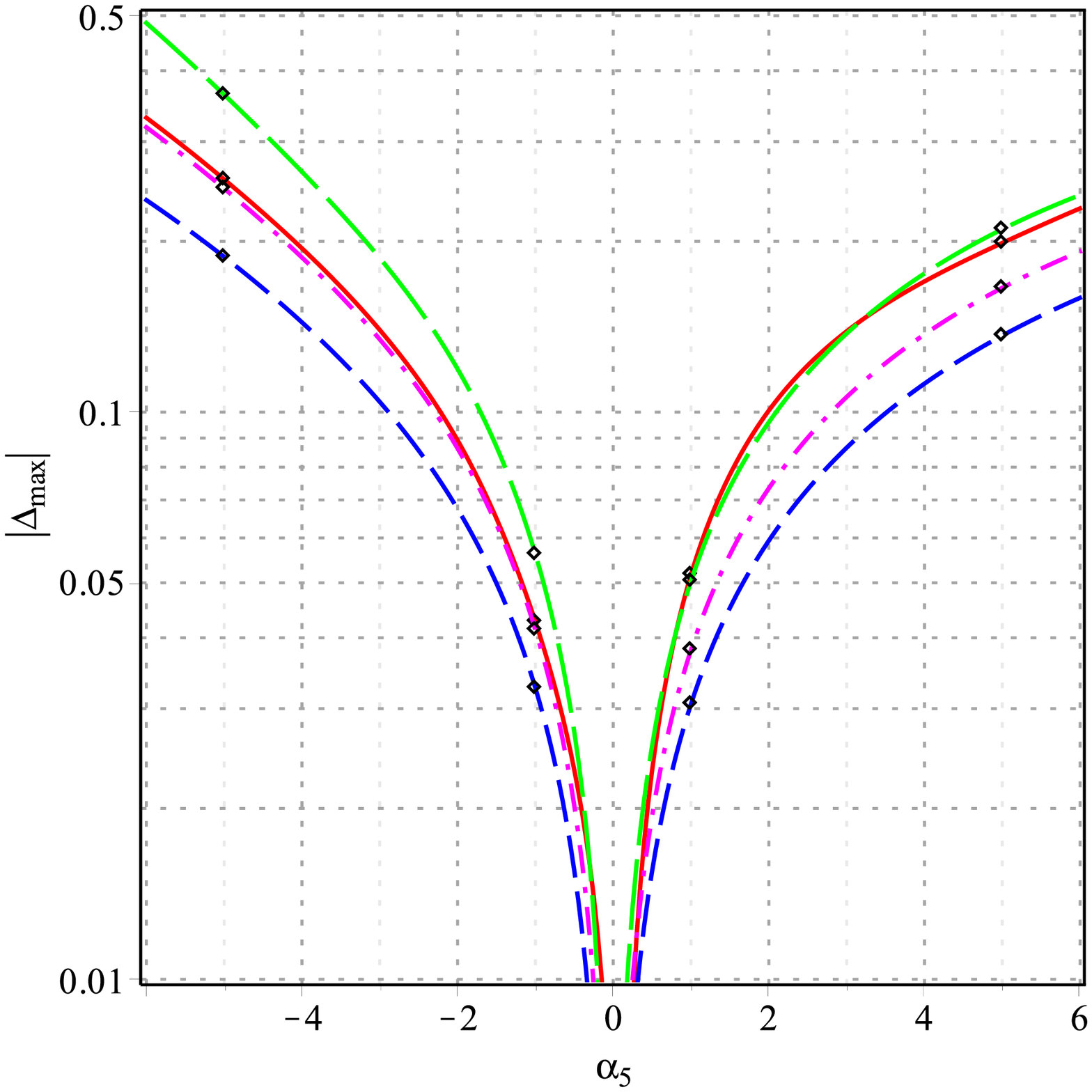,width=0.33\linewidth,clip=} & \epsfig{file=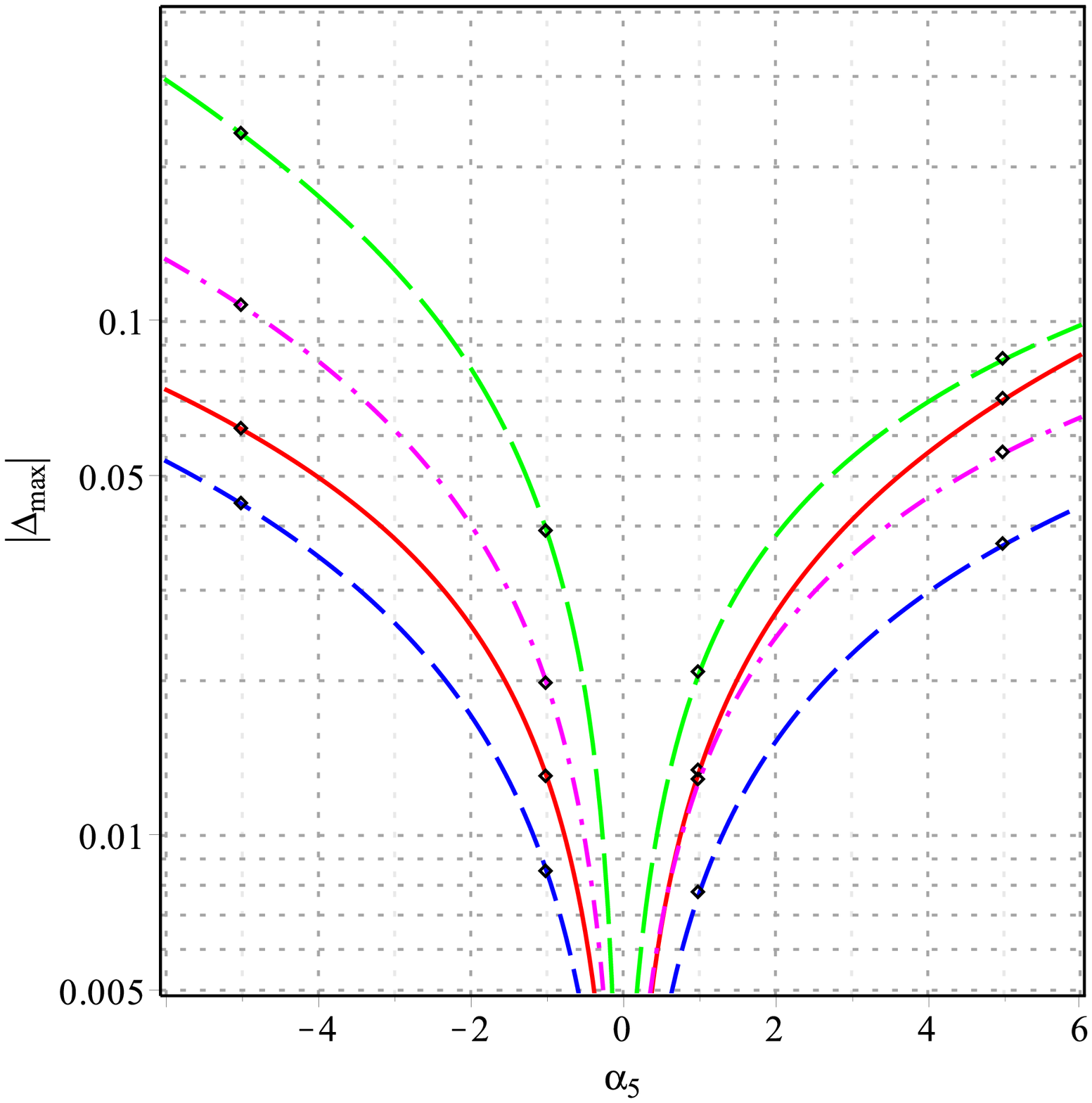,width=0.33\linewidth,clip=} &\epsfig{file=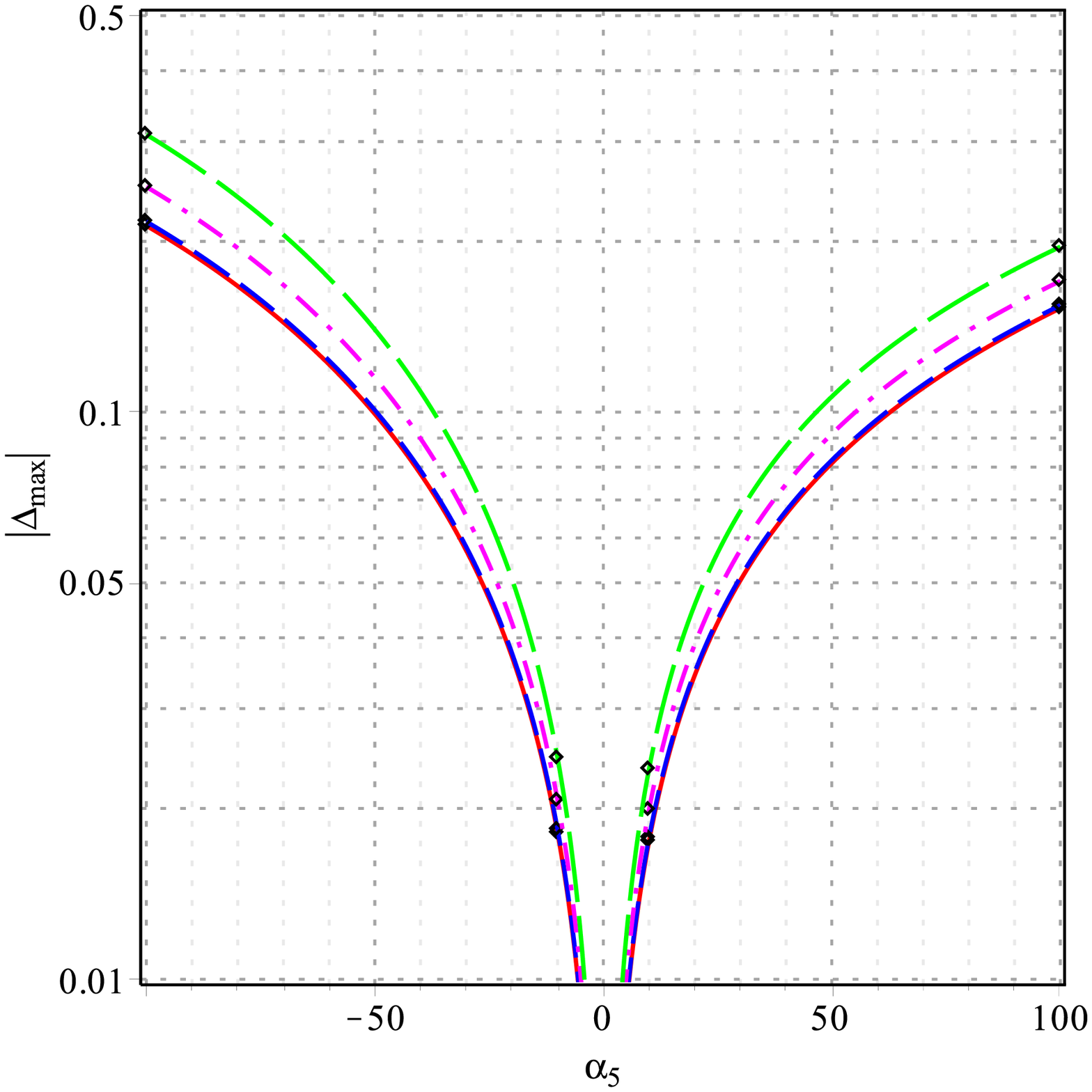,width=0.33\linewidth}\\
\end{tabular}
\caption{(Color online.) The parameter $|\Delta_{\rm max}| = |A_{MG}(r)/A_{GR}(r)-1|_{\rm max}$ as a function of $\alpha_{5}=\alpha/10^5$ for the SLy 
\eos (left), polytropic \eos (middle) and \qs (right). The red (solid), blue (short-dashed), 
magenta (dot-dashed), green (long-dashed) lines indicate
the $\gamma = 0, -0.01, -0.1, -0.25$ cases respectively. A necessary condition for the 
validity of the perturbative approach is $|\Delta_{\rm max}|<1$.  The circles indicate the parameter values used in Figs. \ref{PolyMRdiag}, 
\ref{SLyMRdiag} and \ref{QSMRdiag}. 
}\label{pertplot}
\end{figure}

This quantity varies as a function of radius for each 
star, and also depends on the corresponding central density. 
In Fig. \ref{pertplot},  we have plotted the quantity $|\Delta_{max}|$ as a function of $\alpha_{5}=\alpha/10^5$
(where the subscript ${\rm max}$ refers to the maximum value for a given choice of parameters)
 for the SLy, polytropic and \qs equations of state.

A necessary condition for the 
validity of the perturbative approach is $|\Delta_{\rm max}|<1$. The plots for the SLy and polytropic equations of state (left and middle) 
show that
 the $f(R) = R+\alpha R^2+\beta R^2\ln R/\mu^2$ model  can be treated 
pertubatively for $|\alpha|\lesssim 10^6$.
The dependence of $|\Delta_{\rm max}|$ on $\alpha$ is linear, with the slope depending on the value of $\gamma$. 
Including a small logarithmic term ($\gamma=-0.01$) decreases $|\Delta_{\rm max}|$, however, increasing \g further leads to larger deviations from GR and thus larger values of $|\Delta_{\rm max}|$.
As mentioned above, in the \qs case, we can reach larger values 
of $\alpha$ respect to \ns while remaining in the the perturbative regime.

\section{Summary}\label{summary}

In this article we have considered the effect of a
logarithmic $f(R)$ theory, $f(R)= R+\alpha R^2+\beta R^2 \ln(R/\mu^2)$,
motivated by the form of the one-loop effective action arising from
gluons in curved spacetime, on the structure of relativistic stars.
Unlike many $f(R)$ theories in the literature,
the modifications to General Relativity are significant in the
strong-field regime, which is less well constrained by observations.
Considering the motivation, we treat the model as an effective theory,
valid in
the interior and near vicinity of neutron stars,
where QCD effects play an important role.

An $f(R)$ theory inevitably introduces a scalar degree of freedom,
and in Section \ref{SubSec:Consistency} we have derived the constraints imposed upon the
parameters of the model due to stability and internal consistency requirements.
Unlike the related $R+\alpha R^2$ model, we find that,
when the logarithmic term is a subdominant correction ---
i.e. $|\gamma| = |\beta/\alpha| \ll 1$, which
we assume throughout this work  ---
one can consider positive and negative values of  $\alpha$.
In addition, in the absence of matter, the existence of a stable minimum at $R=0$ forces us to
work with negative values of the coefficient of the logarithmic term $\beta$.

In Section \ref{SubSec:Observations}, we have also considered the constraints imposed upon the model
by observations; in particular relating to the possibility of a fifth force due to the scalar
degree of freedom.
Since we treat the model as an effective theory valid only in the vicinity of ultra-dense
matter, we do not need to contend with cosmological or terrestrial constraints, however,
it is important to consider the effect of the modification on binary pulsars and
direct observations of neutron stars.
Transforming the theory to the Einstein frame, we have shown that the
model exhibits a chameleon effect, completely suppressing the effect of the modification on scales
exceeding a few radii, so that any effect on the orbital motion of a binary system is completely negligible. 
We showed that this model satisfies the binary star observations of the effective 
gravitational constant for a wide range of parameters \a and \g. 

On smaller scales, near the surface of the neutron star, the deviation from General Relativity can be
significant.
Observations of bursting neutron stars depend strongly on the surface redshift $z_s$, which
determines the shift in absorption (or emission) lines due to elements in the atmosphere, as well as the Eddington critical luminosity.
In Fig. \ref{redshift.fig} we have plotted the dependence of $z_s$ on the radial coordinate in
the immediate vicinity of the neutron star surface
(which is directly related to the observable quantity $\delta\lambda/\lambda=z_s$)
showing that there are strong $\alpha$-dependent deviations from General Relativity,
which could in principle be detected, utilising data from future X-ray missions.

In section \ref{NS}, we have
used the method of perturbative constraints to
derive and solve the modified Tolman-Oppenheimer-Volkov equations
for neutron and quark stars.
The changes to the mass-radius diagram for neutron stars are shown in Fig. \ref{PolyMRdiag} for a toy polytropic equation of state and in Fig. \ref{SLyMRdiag} for 
a realistic SLy equation of state. 
As in the $f(R)=R+\alpha R^2$ model \cite{Arapoglu,Orellana:2013gn}
there is an inversion of the modified gravity effect near the  central density
$\rho\simeq 5\rho_{ns}$ for the SLy equation of state. For the SLy equation of state, the deviation from GR is more evident for smaller central densities 
(corresponding to the lower-right of the plots in Fig. \ref{SLyMRdiag}).
However, in the polytropic case, 
for higher central densities (top-left part of the plots in Fig. \ref{PolyMRdiag}) one can observe  a larger
deviation from GR with respect to lower central densities (bottom-right on the plots). In addition, 
in the polytropic case, the deviation from GR is much larger than the SLy case for  
equal values of the parameter \a. For the polytropic equation of state, the asymmetry
in the M-R diagram for positive and negative values of parameter \a is also reduced.

As has been noted in the case of other
$f(R)$ models, there is a degeneracy with the choice of equation of state that is largely unconstrained. 
To break this degeneracy, one could consider other observables, such as those relating 
to the cooling \cite{Page:2005fq} or spin properties \cite{Baubock:2013gna} of the neutron stars. 
In particular, it was suggested in \cite{cooney} that since
cooling by neutron emission --- which is the dominant cooling mechanism for young ($\lesssim 10^4-10^6$ years) 
neutron stars --- is particularly sensitive to the central density of the star, 
measurements of the surface temperature could offer a discriminant. However, in practice, 
the neutrino cooling rate is difficult to model due to the strong dependence
on features such as condensates in the star's composition.

We find that the range of the parameter $\alpha\lesssim10^{6}{\rm m^2}$ that is consistent with the perturbative treatment in our model for the SLy and polytropic equations of state is
comparable with that in related works,
 where $\alpha<10^{9}{\rm cm^2}$ \cite{Arapoglu,Cheoun:2013tsa}, $\alpha\lesssim10^5\; {\rm m^2}$ \cite{Orellana:2013gn}. 
 In the quark star case, one can reach larger values 
of $\alpha \sim 10^7 {\rm m^2}$ while remaining in the the perturbative regime.

Finally, in section \ref{Sec:QuarkStars}, we have considered the case of
self-bound stars, consisting of strange quark matter.
We found that the $M$-$R$ diagram and internal density distribution were insensitive to
the presence of the logarithmic term, and for positive $\alpha$ the mass is always enhanced
relative to that calculated using General Relativity.

As the modified Tolman-Oppenheimer-Volkov equations for the \f model considered here involve $\ln(R/\mu^2)$ terms
that are not well defined at $R=0$ we have restricted our analysis to 
the $R>0$ domain. Since neutron star equations of state are stiff
and  $R\propto(\rho-3P)$, when $\gamma \ne 0$ we cannot consider central densities above a maximum value.
This is particularly evident in Fig. \ref{SLyMRdiag}, as the largest deviations from GR occur 
for stars with low masses, corresponding to a medium central density. Using an equation of state
that is less stiff for large densities would give rise to more significant deviations for larger mass stars. 
This can be seen in the quark star case.

To conclude, we have shown that considering the
finite logarithmic terms arising in the calculation of the effective action
for a gauge field in a phenomenological $f(R)$ framework leads to
interesting observational consequences differing from the
predictions of General Relativity.
To make this connection more definite is beyond the scope of this article,
although as observational data improve,
one can entertain the possibility that neutron star systems
may in the future have a role to play in analysing the
predictions of quantum field theory
in curved spacetime.

\section*{Acknowledgements} 
We would like to thank F. R. Klinkhamer, C. Rahmede, 
V. Emelyanov, J. Creighton and Y. Ek\c{s}i for helpful and useful discussions 
and comments. The work of JMW is supported
 by the ``Helmholtz Alliance for Astroparticle Physics HAP",  funded by the Initiative and Networking Fund of the Helmholtz Association.

\section*{References}

\renewcommand*{\bibfont}{\raggedright}
\bibliographystyle{apsrev4-1}
\bibliography{refer1} 
\end{document}